\documentclass[journal=ancac3,manuscript=article]{achemso}
\setkeys{acs}{usetitle=true}
\setkeys{acs}{maxauthors=20}
\usepackage[version=3]{mhchem} 

\usepackage{xcolor}

\author{Shawulienu Kezilebieke}
\affiliation{Department of Applied Physics, Aalto University School of Science, 00076 Aalto, Finland}
\author{Md Nurul Huda}
\affiliation{Department of Applied Physics, Aalto University School of Science, 00076 Aalto, Finland}
\author{Paul Dreher}
\affiliation{Donostia International Physics Center (DIPC), Paseo Manuel de Lardiz\'abal 4, 20018 San Sebasti\'an, Spain}
\alsoaffiliation{Centro de F\'isica de Materiales (CSIC-UPV-EHU), Manuel Lardiz\'abal 4, 20018 San Sebasti\'an, Spain}
\author{Ilkka Manninen}
\affiliation{Department of Applied Physics, Aalto University School of Science, 00076 Aalto, Finland}
\author{Yifan Zhou}
\affiliation{Department of Applied Physics, Aalto University School of Science, 00076 Aalto, Finland}
\author{Jani Sainio}
\affiliation{Department of Applied Physics, Aalto University School of Science, 00076 Aalto, Finland}
\author{Rhodri Mansell}
\affiliation{Department of Applied Physics, Aalto University School of Science, 00076 Aalto, Finland}
\author{Miguel M. Ugeda}
\affiliation{Donostia International Physics Center (DIPC), Paseo Manuel de Lardiz\'abal 4, 20018 San Sebasti\'an, Spain}
\alsoaffiliation{Centro de F\'isica de Materiales (CSIC-UPV-EHU), Manuel Lardiz\'abal 4, 20018 San Sebasti\'an, Spain}
\alsoaffiliation{Ikerbasque, Basque Foundation for Science, 48013 Bilbao, Spain}
\author{Sebastiaan van Dijken}
\affiliation{Department of Applied Physics, Aalto University School of Science, 00076 Aalto, Finland}
\author{Hannu-Pekka Komsa}
\affiliation{Department of Applied Physics, Aalto University School of Science, 00076 Aalto, Finland}
\author{Peter Liljeroth}
\email{peter.liljeroth@aalto.fi}
\affiliation{Department of Applied Physics, Aalto University School of Science, 00076 Aalto, Finland}

\title{Electronic and Magnetic Characterization of Epitaxial VSe$_2$ Monolayers on Superconducting NbSe$_2$}


\begin{document}
\begin{abstract}
Vertical integration of two-dimensional (2D) van der Waals (vdW) materials with different quantum ground states is predicted to lead to novel electronic properties that are not found in the constituent layers. Here, we present the direct synthesis of superconductor-magnet hybrid heterostructures by combining superconducting niobium diselenide (NbSe$_2$) with the monolayer (ML) vanadium diselenide (VSe$_2$). More significantly, the \emph{in-situ} growth in ultra-high vacuum (UHV) allows to produce a clean and an atomically sharp  interfaces. Combining different characterization techniques and density-functional theory (DFT) calculations, we investigate the electronic and magnetic properties of VSe$_2$ on NbSe$_2$. Low temperature scanning tunneling microscopy (STM) measurements demonstrate a reduction of the superconducting gap on VSe$_2$ layer. This together with the lack of charge density wave signatures indicates magnetization of the sheet, at least on the local scale. However, overall, VSe$_2$ does not behave as a conventional ferromagnet.
\end{abstract}

\newpage
There has been a surge of interest in designer materials that would realize electronic responses not found in naturally occurring materials. There are many routes towards this goal and they are all being explored vigorously: \emph{e.g.}, artificial atomic lattices\cite{Gomes2012DesignerGraphene,Slot2017ExperimentalLattice,Drost2017,Khajetoorians2019designer,Yan2019engineered}, atomically precise graphene nanoribbons\cite{Talirz2016On-SurfaceNanoribbons,Groning2018EngineeringNanoribbons,Rizzo2018TopologicalNanoribbons}, and controlled heterostructures of two-dimensional materials \cite{Geim2013,Novoselov2016_review,Huang2017_CrI3,Gong2017_Cr2Ge2Te6,Bonilla2018,Cao2018UnconventionalSuperlattices,Gibertini2019}. The designer concept is well illustrated in systems combining magnetism with superconductivity to realize topological superconductivity \cite{Zhang2011_RMP,Mourik2012science,Nadj-Perge2014,Sato2017_review,Zhou2019}. Individual magnetic impurity atoms give rise to so-called Yu-Shiba-Rusinov states \cite{Heinrich2018_review}, which can be coupled in extended structures to give rise to bands (inside the superconducting gap). Eventually, the system can be driven into a topological phase in the presence of certain spin textures or spin-orbit coupling \cite{Nadj-Perge2014,Ojanen2014_prb,Kim2018,Roentynen2015,Li2016,Rachel2017}. Topological superconductors are a distinct form of matter that is predicted to host boundary Majorana fermions. Experimental realization of Majorana fermions is exciting in its own right, but this is compounded by the proposal that systems with non-abelian statistics can serve as the basis for topological quantum computation\cite{Nayak2008,Alicea2012,Wilczek2009}. Experimentally, these systems have been realized in one-dimensional chains of magnetic adatoms on the surface of s-wave superconductors\cite{Nadj-Perge2014,Kim2018,Ruby2015} and this has been extended to two-dimensional systems by using magnetic metal islands \cite{Menard2017_NatComm,palacio}. However, these types of systems can be sensitive to disorder and interface engineering through, \emph{e.g.}, the use of an atomically thin separation layer, might be required \cite{palacio}.

Issues with interface inhomogeneities can potentially be avoided in van der Waals (vdW) heterostructures, where the different layers interact only through vdW forces \cite{Geim2013,Novoselov2016_review}. Layered materials that remain magnetic down to the monolayer (ML) limit have been recently demonstrated \cite{Huang2017_CrI3,Gong2017_Cr2Ge2Te6}. While the first reports relied on mechanical exfoliation for the sample preparation, related materials CrBr$_3$ and Fe$_3$GeTe$_2$ have also been grown using molecular-beam epitaxy (MBE) in ultra-high vacuum (UHV) \cite{Liu2017_Fe3Gete2,chen2019direct}. This is essential for realizing clean edges and interfaces. Very recently, monolayer magnetism was suggested in the transition metal dichalcogenide (TMD) vanadium diselenide (VSe$_2$), which can be readily grown using MBE on various layered materials \cite{Bonilla2018}. Later reports have questioned the existence of magnetism in VSe$_2$ as no magnetic signal was detected in X-ray magnetic circular dichroism (XMCD) experiments \cite{Feng2018_NanoLett,Wong2019_frustrated}. Angle-resolved photoemission spectroscopy revealed an enhanced charge-density wave (CDW) transition at a higher temperature than in the bulk, and it was suggested that the presence of CDW driven by Fermi-surface nesting removes the usual mechanism for achieving a magnetic ground state \cite{Chen2018_PRL,Duvjir2018_NanoLett,Feng2018_NanoLett,Coelho_2019jpcc}. It has also been suggested that spin frustration plays a role in VSe$_2$ samples grown by MBE \cite{Wong2019_frustrated}. Support for this comes from recent experiments on chemically exfoliated VSe$_2$ flakes that show ferromagnetic behaviour with a Curie temperature of \emph{ca.}~470 K and also non-zero XMCD signal \cite{Yu2019chemically}. While there still is no consensus on the nature of the possible magnetic ground state of VSe$_2$, it is clear that there is a delicate balance between different competing interacting states and phases in monolayer transition-metal dichalcogenides, which may also depend on the nature of the substrate.\cite{Feng2018_NanoLett,Coelho_2019jpcc,Wong2019_frustrated,Sugawara2019_PRB,Kim2019_arXiv,Fumega2019,Yu2019chemically} 

Combining 2D magnetic and superconducting TMDs would create a promising material platform for investigating the coexistence of superconductivity, magnetism and the resulting emergent quantum phases of matter. The inherent lack of surface bonding sites due to the layered nature of these materials prevents chemical bonding between the layers and results in a better control of the interfaces. We report growth of single layer vanadium diselenide (VSe$_2$) by molecular beam epitaxy on superconducting niobium diselenide (NbSe$_2$) substrate and study the magnetic and electrical properties of these heterostructures. MBE growth under UHV conditions facilitates the formation of clean edges and interfaces and we characterize the electronic structure of the resulting hybrid materials down to the atomic scale using low-temperature scanning tunneling microscopy (STM) and spectroscopy (STS). Our results give further experimental information on the magnetic properties of VSe$_2$ and demonstrate a clean and controllable platform for creating superconducting-magnetic hybrid TMD materials with great potential of integrating TMDs into future electronics devices.

\section{Results and discussion}
\begin{figure}[!b]
	\centering
		\includegraphics [width=0.95\textwidth] {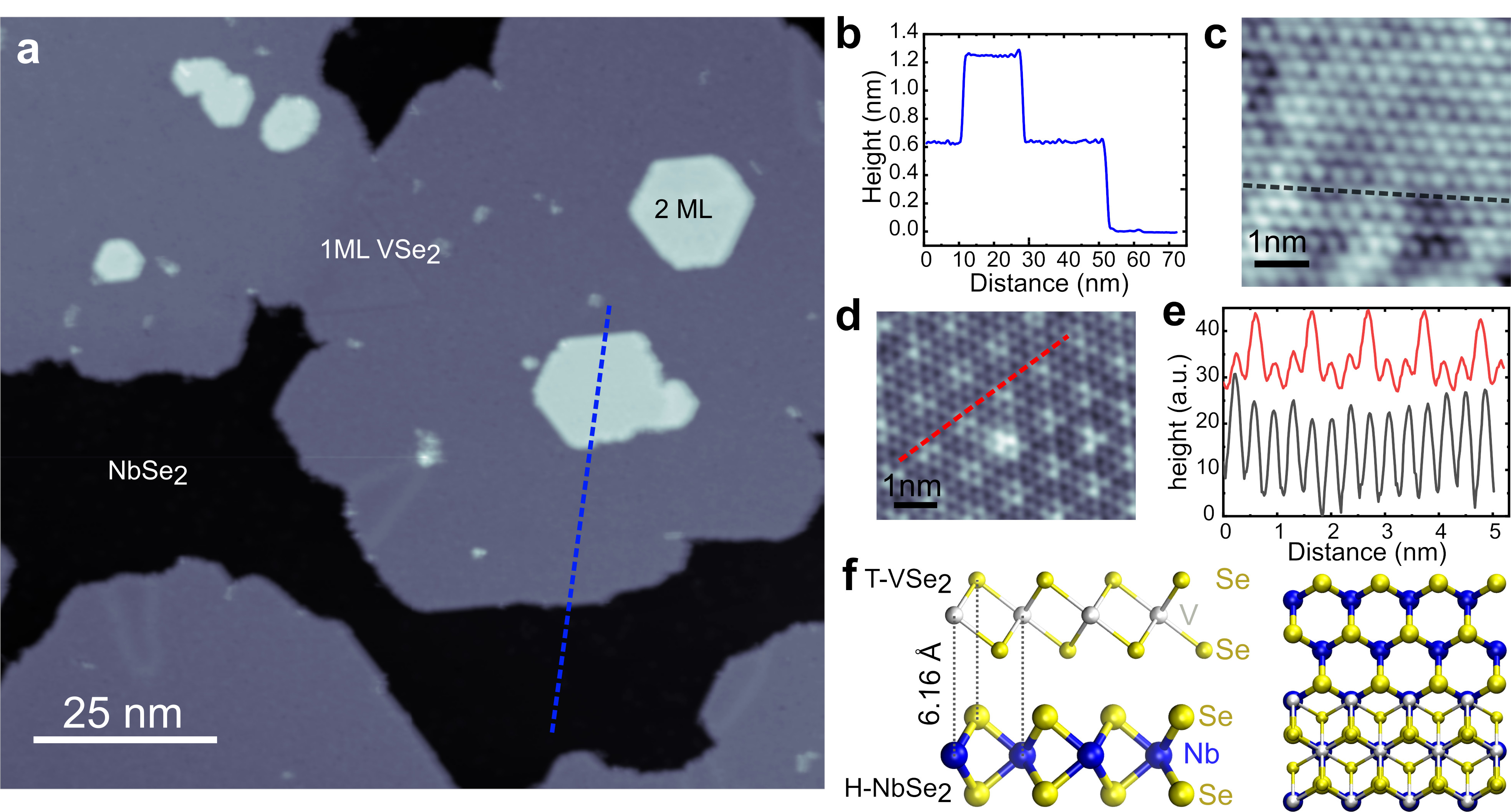}
	\caption{Growth of VSe$_2$ on NbSe$_2$. (a) Large scale STM image of submonolayer VSe$_2$ on NbSe$_2$. (b) Line-profile along the blue line shown in panel a. (c,d) Atomically resolved images on VSe$_2$ (c) and NbSe$_2$ (d). (e) Line-profiles along the lines in panels c and d (VSe$_2$ (black), NbSe$_2$ (red)) showing the atomic periodicities and the charge-density wave modulation on the NbSe$_2$ substrate. (f) Computed structure of VSe$_2$ on NbSe$_2$. }
	\label{fig1}
\end{figure}
VSe$_2$ was grown on NbSe$_2$ by MBE and results are illustrated in Fig.~\ref{fig1} (see Methods for details). Briefly, vanadium was evaporated under excess flux of selenium onto a NbSe$_2$ crystal cleaved \emph{in-situ} in UHV and held at $T = 520-540$ K during the growth. The samples were characterized \emph{in-situ} by STM and X-ray photoelectron spectroscopy (XPS). In addition, after capping the films with a thick Se layer, the samples were characterized \emph{ex-situ} by temperature dependent magnetization measurements. Fig.~\ref{fig1}a shows STM characterization of sub-monolayer VSe$_2$ films on NbSe$_2$ substrate. VSe$_2$ grows atomically smooth, large uniform ML islands. Higher coverages result in the formation of a second layer. The profile along the blue line in Fig.~\ref{fig1}a shows that the apparent height of the VSe$_2$ film is 6.5 \AA (Fig.~\ref{fig1}b) consistent with the unit cell height \cite{Bayard1976}. Atomically resolved STM images of the VSe$_2$ monolayer and  NbSe$_2$ crystal surface are shown in Figs.~\ref{fig1}c and \ref{fig1}d, respectively. While NbSe$_2$ shows the well-known $3\times3$ charge-density wave modulation in the atomic contrast, we do not detect a charge density wave on VSe$_2$ (even at temperature of $T=4.2$ K). This is in contrast to reports on HOPG and bilayer graphene substrates\cite{Bonilla2018,Feng2018_NanoLett,Chen2018_PRL}. The lattice constants can be measured from the atomically resolved images, as depicted in Fig.~\ref{fig1}e. This yields values of $3.5\pm0.1$ \AA~and $3.4\pm0.1$ \AA~ for VSe$_2$ and NbSe$_2$, respectively. These values match well with previous experimental results\cite{Bayard1976,Duvjir2018_NanoLett,Liu2018}. The lattice mismatch is roughly 3\%; this together with the fact that we observe several different orientations of VSe$_2$ w.r.t.~the underlying NbSe$_2$ suggests that there is no lattice match between VSe$_2$ and NbSe$_2$. 

XPS was used to study the chemical composition of the as-grown VSe$_2$ films on NbSe$_2$. Characteristic peaks of V, Se and Nb are found in the XPS spectra (see Supporting Information (SI) Fig.~S1). The binding energies of the V 2p$_{3/2}$ peak at 513.7 eV and the Se 3d$_{5/2}$ peak at 53.4 eV are similar to those previously observed for VSe$_2$ on HOPG\cite{Liu2018,Bonilla2018}. The Nb 3d$_{5/2}$ peak is found at 203.5 eV which is typical for NbSe$_2$. Both VSe$_2$ and NbSe$_2$ have similar selenium binding energies and thus they cannot be resolved from the Se 3d spectrum \cite{Wang2017}. The V:Se:Nb stoichiometry for a 0.6 ML VSe$_2$ film was found to be roughly 1:5:2 (estimated from integrated peak areas which were normalized to the elements' atomic sensitivity factors). No other elements, such as possible magnetic impurities, were detected above the detection limit of $\sim$1 atomic percent.

We have complemented the experiments by density functional theory (DFT) calculations (see Methods and SI for details). Fig.~\ref{fig1}f shows the fully relaxed geometry of VSe$_2$/NbSe$_2$ heterostructure from the side and top views. The energetically most favourable stacking has the lower layer Se atoms of VSe$_2$ on top of hollow site of the NbSe$_2$ (2.94 \AA~from Se in NbSe$_2$ to Se in VSe$_2$) and V on top of Nb, with a distance of 6.16 \AA~from Nb to V.

\begin{figure}[!t]
	\centering
		\includegraphics [width=0.5\textwidth] {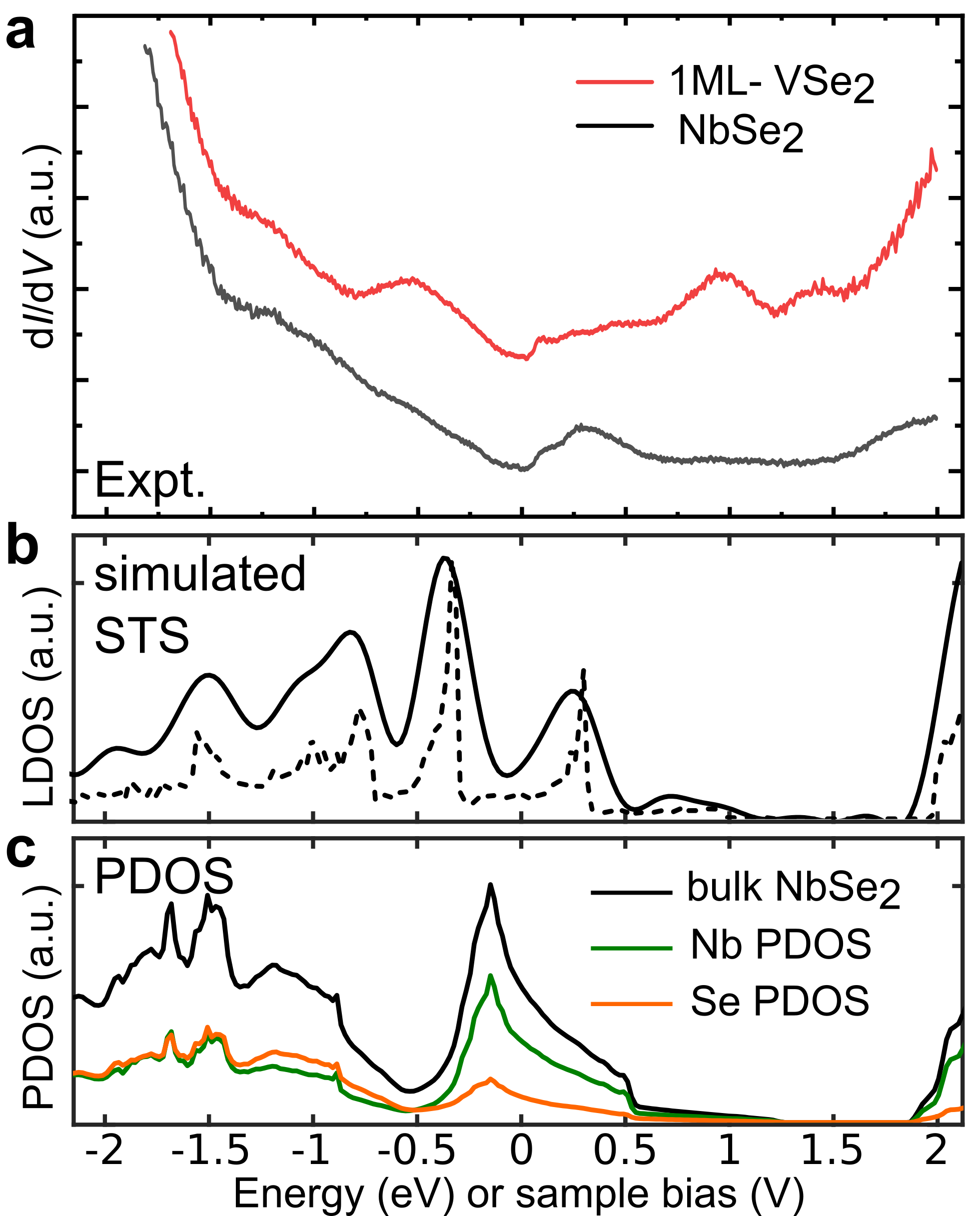}
	\caption{(a) Typical long-range experimental d$I$/d$V$ spectra on a ML VSe$_2$ and on the NbSe$_2$ substrate. (b) Simulated STS obtained by integrating LDOS at a constant height on top of the NbSe$_2$ surface. The solid and dashed lines show different values of the energy broadening. (c) Calculated density of states of bulk NbSe$_2$. Density of state plots show both the total DOS and the DOS projected to the metal and Se atoms. }
	\label{fig2}
\end{figure}
We have probed the electronic structure of single-layer VSe$_2$/NbSe$_2$ heterostructure by both scanning tunnelling spectroscopy (STS) and DFT. Fig.~\ref{fig2}a shows typical d$I$/d$V$ spectra taken on the ML VSe$_2$ and on the bulk NbSe$_2$ substrate over a large bias range. We will first focus on the NbSe$_2$ response. At positive bias (empty states) region, the most pronounced features on bulk NbSe$_2$ are the broad resonances at $\sim0.3$ V and $\sim1.8$ V, while at negative bias the d$I$/d$V$ signal is broad and rising. The measured d$I$/d$V$ spectrum on NbSe$_2$ is in agreement with earlier STS studies on bulk NbSe$_2$\cite{Arguello2014}. The features in our d$I$/d$V$ spectroscopy also match the simulated spectra on a 3-layer slab of NbSe$_2$ (Fig.~\ref{fig2}b) and can be compared with the bulk density of states (Fig.~\ref{fig2}c). The first resonance at positive bias arises from the Nb-derived band, while the broad feature at negative bias overlaps with the mostly selenium derived bands below $E_\mathrm{F}$. On VSe$_2$, at positive bias, there are pronounced features close to the Fermi level and also at 0.9 V and 1.45 V. At negative bias, we observe a peak at -0.5 V, and a shallow feature at $\sim -1$ V. These VSe$_2$ features are discussed in more detail below.

\begin{figure}[!t]
	\centering
		\includegraphics [width=0.95\textwidth] {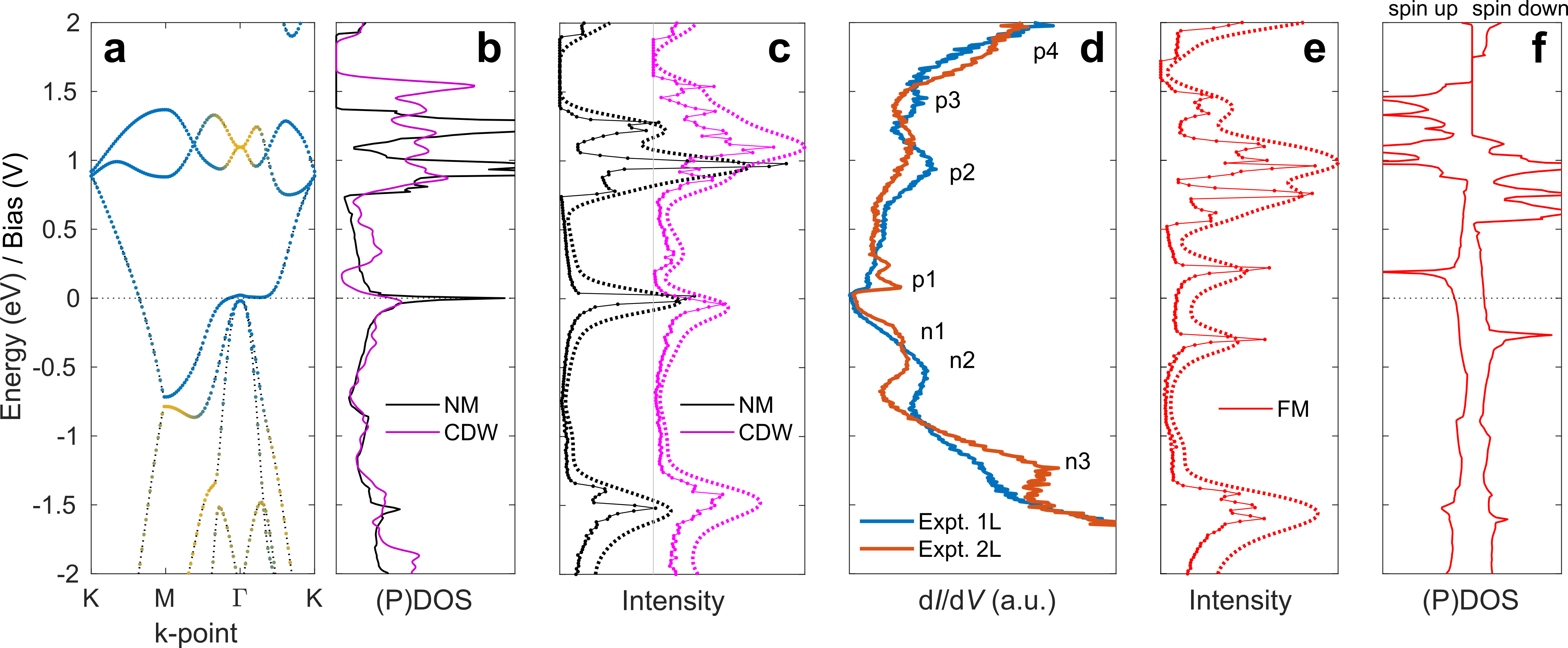}
	\caption{(a) Calculated band structure of VSe$_2$ (non-magnetic state). (b,c) Calculated PDOS (b) and simulated d$I$/d$V$ spectra (with two different broadenings, panel c) for the non-magnetic and charge-density wave states. (d) Typical long-range experimental d$I$/d$V$ spectra on 1 and 2 ML VSe$_2$ on NbSe$_2$. (e,f) Simulated d$I$/d$V$ spectra (with two different broadenings, panel e) and calculated PDOS (f) for the ferromagnetic ground state. }
	\label{fig3}
\end{figure}

The electronic structure and the simulated and experimental STS are shown in Fig.~\ref{fig3}. First focusing on the experimental spectra for monolayer and bilayer VSe$_2$ in Fig.~\ref{fig3}d,
we observe peaks close to the Fermi level at both positive and negative bias and they are more pronounced for the bilayer compared to the monolayer. Their energy spacing (gap) is $\sim 0.2$ eV with an abrupt edge at positive energy (peak p1) and smoother edge at negative energy that develops to peaks labelled with n1/n2. At larger positive or negative bias, several peaks can be distinguished with the peak positions shifting between the mono- and bilayer spectra.

Starting from the non-magnetic (NM) phase, the band structure shows the partially filled d-band (Fig.~\ref{fig3}a). There is a flat region between $\Gamma$ and K-points, which also happens to coincide with the Fermi-level. This leads to a strong peak at the Fermi-level in the DOS and also in the simulated STS, which obviously contrasts the experimental spectra. NM phase is also unstable in calculations and may either develop CDW or ferromagnetism, depending on conditions (strain, doping, defects) and on computational parameters (see SI). In the CDW phase, DOS exhibits a clear pseudogap formation, although not at the Fermi-level, but just above it. While the simulated STS would be relatively consistent with the experimental one,  we do not detect CDW in the STM topography and also do not observe a hard gap at the Fermi-level as reported in Ref.~\citenum{Bonilla2018}. 

In the ferromagnetic (FM) phase, the spin-up and -down band structures are only shifted in energy as in itinerant FM. The calculated magnetic moment depends moderately on strain and the computational parameters (\emph{e.g.}~the Hubbard $U$), see SI for details \cite{Ma2012}. Consequently also the DOS and simulated STS show splitting of the peaks w.r.t.~NM phase. Due to the splitting of the DOS peak at Fermi-level, the lower branch could explain the peaks n1/n2 with the higher branch falling slightly above Fermi-level (feature p1 in the experimental spectra). The FM phase also seems to yield better agreement between the simulated and experimental STS for the peaks p2 and p3 in the monolayer sample. Even though the magnetic response of VSe$_2$ is rather complex, at the atomic scale features in the STS spectra seem to have better match with the FM phase than the CDW phase.

\begin{figure}[!t]
	\centering
		\includegraphics [width=0.5\textwidth] {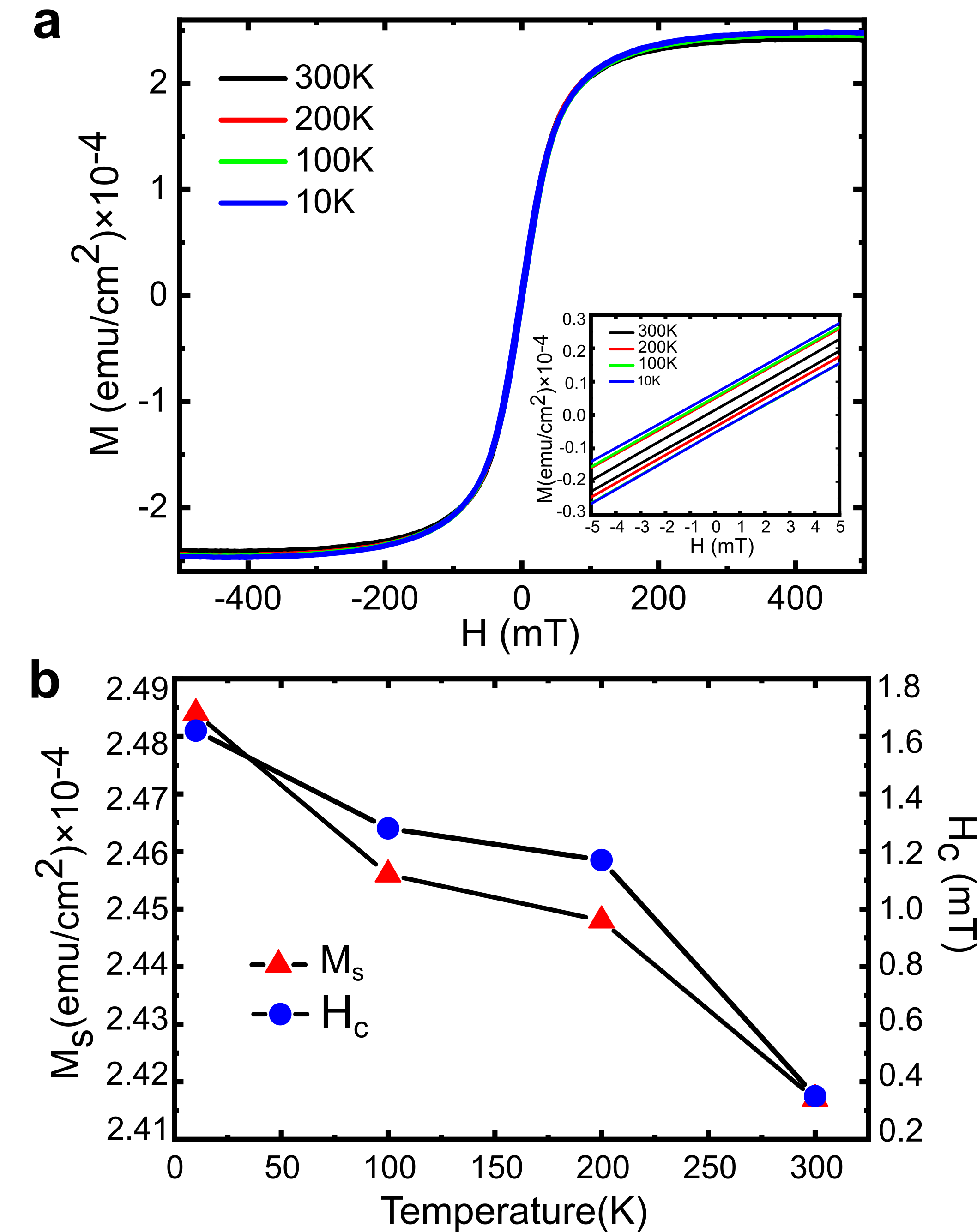}
	\caption{Magnetic measurements on ML VSe$_2$ on NbSe$_2$. (a) Background corrected magnetization curves loops taken at $T=10-300$ K. (b) The temperature dependencies of $M_\mathrm{s}$ and $H_\mathrm{c}$.}
	\label{fig4}
\end{figure}

After the electronic characterization of the samples, we will next focus on their magnetic properties. To explore the magnetic properties of ML VSe$_2$ on NbSe$_2$, we carried out magnetization measurements at various sample temperatures (See Methods for details). All the VSe$_2$ samples measured showed an in-plane magnetic response similar to that shown in Fig.~\ref{fig4}a, where the loops have a vanishing coercivity and remanence, but show saturation around $200-300$ mT \cite{Duvjir2018_NanoLett}. A paramagnetic background has been subtracted from this data (see SI Fig.~S14; Fig.~S15 shows a direct comparison between the substrate and VSe$_2$ responses). As shown in Fig.~\ref{fig4}b, the coercive field is very small and it and saturation magnetization are practically independent of temperature in the range of $T=10-300$ K. While the mechanism is different here, there are other systems that exhibit phenomenologically similar magnetic responses \cite{Coey_2016,Coey2019}. This behaviour is inconsistent with standard ferromagnetism and it has been suggested that spin frustration plays a role in samples grown by MBE \cite{Wong2019_frustrated,Yu2019chemically}. Experiments on chemically exfoliated VSe$_2$ flakes exhibit ferromagnetic response with a Curie temperature of \emph{ca.}~470 K \cite{Yu2019chemically}. The point on these experiments was that chemical exfoliation from the bulk results in macroscopic single-crystalline flakes of VSe$_2$ that then exhibit the expected magnetic response. As there is no strict alignment between the substrate and the VSe$_2$ layer, MBE growth results in a polycrystalline 2D layer, which also may be affected by substrate interactions. However, local measurements (such as STM) should still probe the ferromagnetic phase that corresponds to the DFT calculations.

It has been argued that there is no conventional ferromagnetism in VSe$_2$ on graphite and graphene substrates\cite{Chen2018_PRL,Duvjir2018_NanoLett,Feng2018_NanoLett,Wong2019_frustrated,Fumega2019}. Here, one potentially important difference between our samples and those on graphite and graphene is the absence of a charge-density wave: we find no evidence of CDW in either the STM images or the d$I$/d$V$ spectroscopy (see below). The stability of the CDW according to our DFT calculations is extensively discussed in the SI. Furthermore, in previous studies \cite{Bonilla2018}, monolayer VSe$_2$ sample exhibits a maximum in $M_\mathrm{s}$ and $H_\mathrm{c}$ in a range around $T=100$ K and this non-monotonic behaviour is ascribed to the CDW transition. This is in contrast to our data on $M_\mathrm{s}$ and $H_\mathrm{c}$ as a function of temperature. The absence of CDW transation is also independently determined from the magnetization measurement under zero-field-cooled (ZFC) and field-cooled (FC) regimes, where we do not observe noticeable changes in the slope of the ZFC and FC curves due to the CDW (see SI Fig.~S16).

The magnetic response of the sample below the superconducting transition temperature ($T_c = 7.2$ K) is dominated by the superconducting NbSe$_2$ substrate. Magnetization $M(H)$ curves measured at 2 K for bulk NbSe$_2$ and VSe$_2$/NbSe$_2$ (see SI Fig.~S16) show superconducting diamagnetic behavior as reported previously\cite{Valle2017,Galvan2003}. Moreover, there is no obvious difference between these magnetization curves, which indicates that the signal is mostly dominated by the bulk NbSe$_2$. This conclusion is further supported by the temperature dependent magnetization curves for the zero-field cooling (ZFC) of a bulk NbSe$_2$ and VSe$_2$/NbSe$_2$ heterostructure (see SI). They show a rapid decrease at the onset of the diamagnetic signal below the critical temperature $T_\mathrm{c}$ for both samples and, again, there is no obvious change in $T_\mathrm{c}$, which further indicates that the dominance of the bulk signals. It is worth to mention that $T_\mathrm{c}$ is very sensitive to the magnetic doping, with studies suggesting that $T_\mathrm{c}$ drops rapidly upon metal atom doping\cite{Naik2013,Pervin2017}. This suggests that we do not lose any vanadium due to intercalation at the normal growth temperatures. Increasing the growth temperature to $T>300$ $^\circ$C results in intercalation of vanadium, which is clearly seen in atomically resolved images of the NbSe$_2$ surface and results in the loss of the long range order of the CDW \cite{Chatterjee2015} (see SI Fig.~S17 for details).

There is a particular interest in the interaction between the superconducting substrate and the magnetic layer both in terms of the proximity effect induced in the single layer VSe$_2$ and conversely, the effect of the magnetic layer on the underlying superconductor. The superconducting proximity effect can be used to spontaneously drive a non-superconducting material (normal metal) into superconductivity, however, this picture is altered when the superconductor makes a contact with a magnetic layer. In the case of a ferromagnet, the superconducting order parameter is expected to decay exponentially with a very short coherence length $ \zeta _F$ (typically some nm) at the superconductor-ferromagnet (SF) interface. Moreover, the pairing potential $\Delta _p$ inside the ferromagnet shows a strong oscillatory and damped behaviour due to the internal exchange field of the ferromagnet \cite{Linder2015,Buzdin2005}. This type of phenomena could also occur in our hybrid VSe$_2$/NbSe$_2$ layers and it will allow us to shed some light on the nature of the magnetism in VSe$_2$ on the atomic scale.
\begin{figure}[!t]
	\centering
		\includegraphics [width=0.95\textwidth] {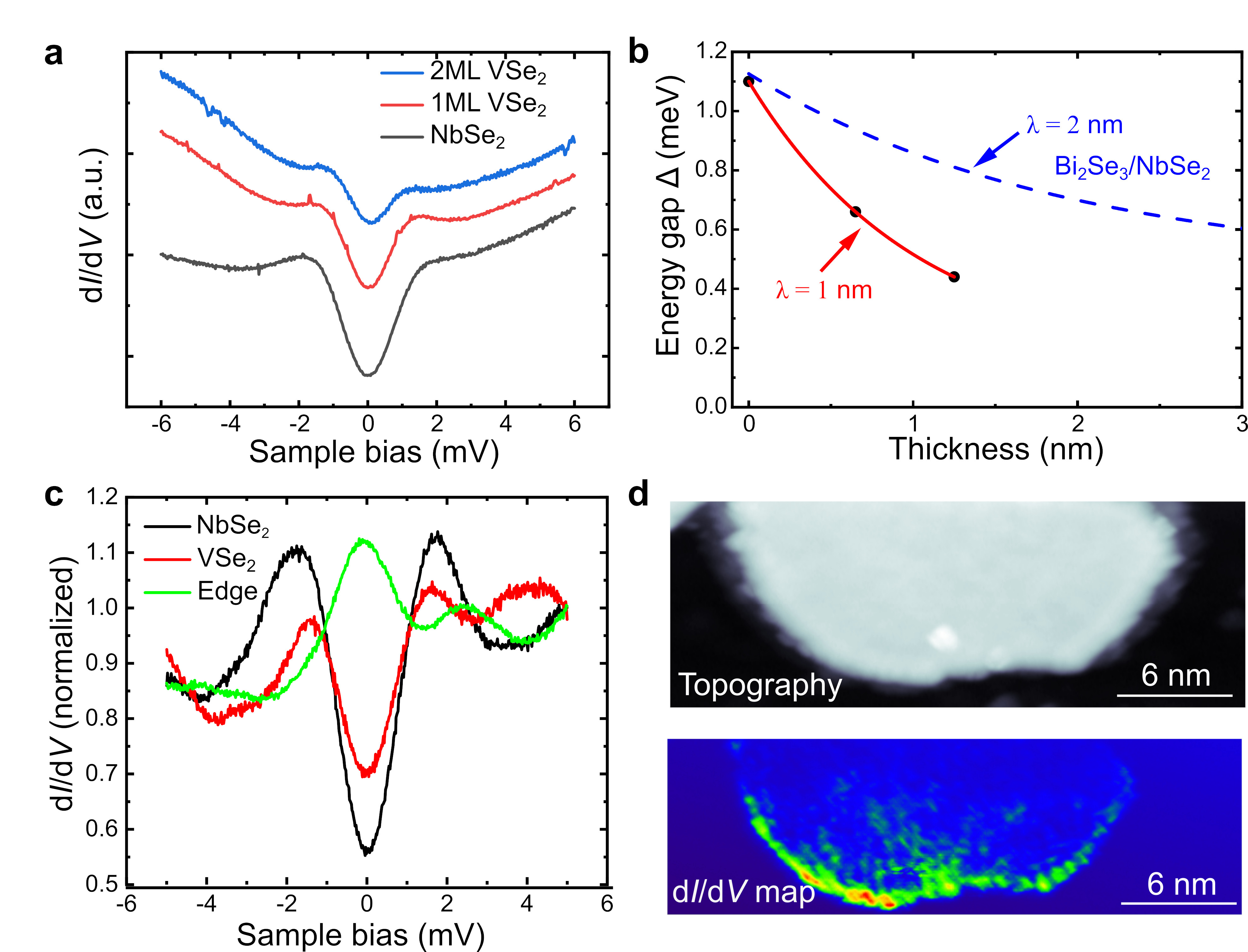}
	\caption{d$I$/d$V$ spectroscopy on VSe$_@$ layers on NbSe$_2$. (a) Spectra measured on clean NbSe$_2$ and on 1 and 2 ML of VSe$_2$ (spectra offset vertically for clarity). (b) The measured energy gap as a function of the VSe$_2$ thickness. (c) Spectra measured at different locations, including over the edge of a VSe$_2$ island. (d) Topographic STM image and a constant-height d$I$/d$V$ map at zero bias over an edge of a VSe$_2$ island.}
	\label{fig5}
\end{figure}

Fig.~\ref{fig5}a shows the d$I$/d$V$ spectra measured on the NbSe$_2$ substrate as well as VSe$_2$ layers with different thicknesses. On NbSe$_2$, we observe a typical superconducting gap: a pronounced dip in the DOS at the Fermi level and coherence peaks on both sides of the gap\cite{hudson1999thesis,Noat2015_nbse2}. The spectra measured on the VSe$_2$ films with thicknesses of 1 ML and 2 ML also shows a superconducting gap, but the gap width is significantly reduced compared to the bare NbSe$_ 2$ substrate. We do not observe oscillatory behaviour of the pairing potential as bilayer VSe$_2$ is not sufficiently thick for this. To further quantify the reduction of the SC gap width, we use a simple analytical relation between a superconducting gap $\Delta$ and  decay length $\lambda$ \cite{Reeg2016,Belzig1996,Floetotto2018}: $\Delta \approx \Delta _\mathrm{NbSe_2} e^{-d /\lambda}$, where the $d$ is the VSe$_2$ film thickness and $\Delta_\mathrm{NbSe_2}$ is the gap of bare NbSe$_2$, respectively. We extract the apparent gap widths from the spectra shown in Fig.~\ref{fig5}a by fitting them (see SI Fig.~S18 for details) and plot the results in Fig.~\ref{fig5}b. The decrease of the energy gap follows an exponential dependence with an decay length of $\lambda = 1$ nm (1 ML thickness is roughly 0.65 nm). This is a measure of the quasi-particle coherence length associated with Andreev reflections. However, this decay length is much shorter than recent experimental results on Bi$_2$Se$_3$ on NbSe$_2$\cite{Wang2012}: they observe a decay length of $\lambda=2$ nm, which is shown as a dashed blue line in Fig.~\ref{fig5}b. In the case of Bi$_2$Se$_3$ on Nb, an even much higher value of $\lambda=8.4$ nm was reported\cite{Floetotto2018}. The faster decay we observe in VSe$_2$ is most easily explained by magnetism of the VSe$_2$ layer. In this case, one would expect a shorter coherence length $ \zeta_F$ governed by the magnetization in the VSe$_2$ layer and not by the diffusion (which is a case in Bi$_2$Se$_3$/NbSe$_2$). 
The reduced gap is very uniform within the VSe$_2$ islands and also between different islands. This shows that the observed response is not affected by the orientation of the VSe$_2$ islands w.r.t. the underlying NbSe$_2$ and that it is related to bulk properties of ML VSe$_2$ and not simply an effect arising from impurities, vacancies, structural imperfections (\emph{e.g.}~layer edges).

We have also probed the spatially dependent spectroscopic response over the edges of the VSe$_2$ islands and typical spectra are shown in Fig.~\ref{fig5}c (more results in the SI Fig.~S19). The spectra evolve from the typical gapped structure over the NbSe$_2$ into a sharp peak at zero bias at the edge of the VSe$_2$ island. This feature is very localized at the edge of the VSe$_2$ layer. Furthermore, it is inhomogeneously distributed along the edges of VSe$_2$ islands and there are strong intensity variations as illustrated in Fig.~\ref{fig5}d. In addition to the spatial distribution of the zero bias peak, its width is also strongly position dependent. We observe both zero bias peaks that are confined within the superconducting gap (\emph{e.g.}~Fig.~\ref{fig5}c), but on some other locations (see SI Fig.~S19), its width can be a couple of times larger than the superconducting gap width. Features inside the superconducting gap could arise from Yu-Shiba-Rusinov bands or topological edge modes, but the broader peaks suggest the presence of free unpaired spins giving rise to the Kondo effect\cite{Franke2011_science,Heinrich2018_review}. In any case, the reason is likely related to the changes of the gap structure of the underlying superconductor due to local magnetic fields arising from the edges of the VSe$_2$ layer. These results suggest importance of edge effects in the magnetism of the VSe$_2$ layer.

\section*{Conclusions}
In conclusion, we have demonstrated high-quality epitaxial growth of VSe$_2$-NbSe$_2$ hybrid structures using MBE. We have observed significant and spatially uniform reduction of the superconducting gap of the NbSe$_2$ substrate on the VSe$_2$ islands with the reduction being thickness dependent and stronger on bilayer VSe$_2$. This would be most naturally explained to result from magnetization of the VSe$_2$ layer. The other electronic and magnetic characterization results are also more consistent with magnetization than with charge density waves. Finally, we observe strongly position-dependent, enhanced d$I$/d$V$ intensity at the Fermi level around the edges of the VSe$_2$ layer suggesting that the atomic-scale structural details of the edge of monolayer VSe$_2$ may contribute to its unusual magnetic response. Finally, our work suggests that it will be possible to combine 2D TMDs with different quantum ground states to stimulate new work in the field of 2D-TMDs hybrids.

\section{Methods}

\textbf{Sample preparation.} Monolayer VSe$_2$  was grown on  NbSe$_2$  substrates by e-beam evaporation of V (99.8 \%, Goodfellow Cambridge Ltd.) and simultaneous deposition of atomic Se (99.99 \%, Sigma-Aldrich) from a Knudsen cell under ultra-high vacuum conditions (UHV, base pressure $\sim 10^{-10}$). The NbSe$_2$  substrate (HQ Graphene) was cleaved in vacuum and subsequently annealed in ultra-high vacuum at $T=600$ K for 1 h before film growth. VSe$_2$  was grown at a substrate temperature of $T=520-540$ K. The growth was carried out under Se-rich conditions. The excess selenium desorbs from the substrate since the substrate temperature was higher than the evaporation temperature of selenium atoms ($T=393$ K). The samples were either characterized \emph{in-situ} by XPS and STM or capped with Se by deposition of Se at room temperature before transferring them out of the UHV system.

\noindent
\textbf{XPS measurements.} XPS measurements (Surface Science Instruments SSX-100 spectrometer) were performed using monochromated Al K-alpha radiation with X-ray power of 200 W, a pass energy of 100 eV and a measurement spot size of 1 mm.

\noindent
\textbf{STM measurements.} The STM experiments (Unisoku USM-1300) of the samples were performed at $T=4$ K. STM images were taken in the constant current mode. d$I$/d$V$ spectra were recorded by standard lock-in detection while sweeping the sample bias in an open feedback loop configuration, with a peak-to-peak bias modulation of 5 mV (long-range spectra) or 0.1 mV (short-range spectra of the superconducting gap) at a frequency of 709 Hz.

\noindent
\textbf{Magnetic characterization.} 
Magnetic characterization was carried out using a Quantum Design Dynacool PPMS operating as a vibrating sample magnetometer (VSM). For magnetization measurements the sample is attached to a quartz rod.

\noindent
\textbf{DFT calculations.} 
All density-functional theory calculations are carried out in the plane-wave basis in the projector augmented wave framework as implemented in VASP \cite{kres1,kres2,PAW}. In all calculations, we use 500 eV cutoff and $k$-point sampling corresponding to $24\times24$ mesh in the primitive cell. High $k$-point mesh is required to correctly describe \emph{e.g.} the CDW phases. Further computational details are given in the SI.

V tends to exhibit strong Coulomb correlations, which usually necessitates using either hybrid functionals or +U. It was shown in Ref.\ \citep{Fumega2019}, that depending on the U-parameter, the monolayer is either in nonmagnetic CDW state or in ferromagnetic no-CDW state. This balance is also affected by the number of electrons \cite{Fumega2019} and likely also by strain. Further details on the computational results on the interplay between magnetism and CDW stability is given in the SI.

\begin{acknowledgement}
	This research made use of the Aalto Nanomicroscopy Center (Aalto NMC) facilities and was supported by the European Research Council (ERC-2017-AdG no.~788185 ``Artificial Designer Materials'') and Academy of Finland (Academy professor funding no.~318995 and 320555, Academy postdoctoral researcher no.~309975, and Academy research fellow no.~311058). Our DFT calculations were performed using computer resources within the Aalto University School of Science ``Science-IT'' project and the Finnish CSC-IT Center for Science. M.M.U.~acknowledges support by the Spanish MINECO under grant no.~MAT2017-82074-ERC and by the ERC Starting grant LINKSPM (Grant 758558).

\end{acknowledgement}

\begin{suppinfo}
The Supporting Information is available free of charge on the ACS Publications website.
\begin{itemize}
    \item Computational details.
    \item Further magnetization, XPS and STM experiments.
    \item Effect of the substrate temperature on vanadium intercalation during VSe$_2$ growth.
    
\end{itemize}

\end{suppinfo}

\bibliography{vse2}


\end{document}


\newpage
\section{X-ray photoelectron spectra from as-grown VSe$_2$ on NbSe$_2$}
\begin{figure}[!h]
	\centering
		\includegraphics [width=1\textwidth] {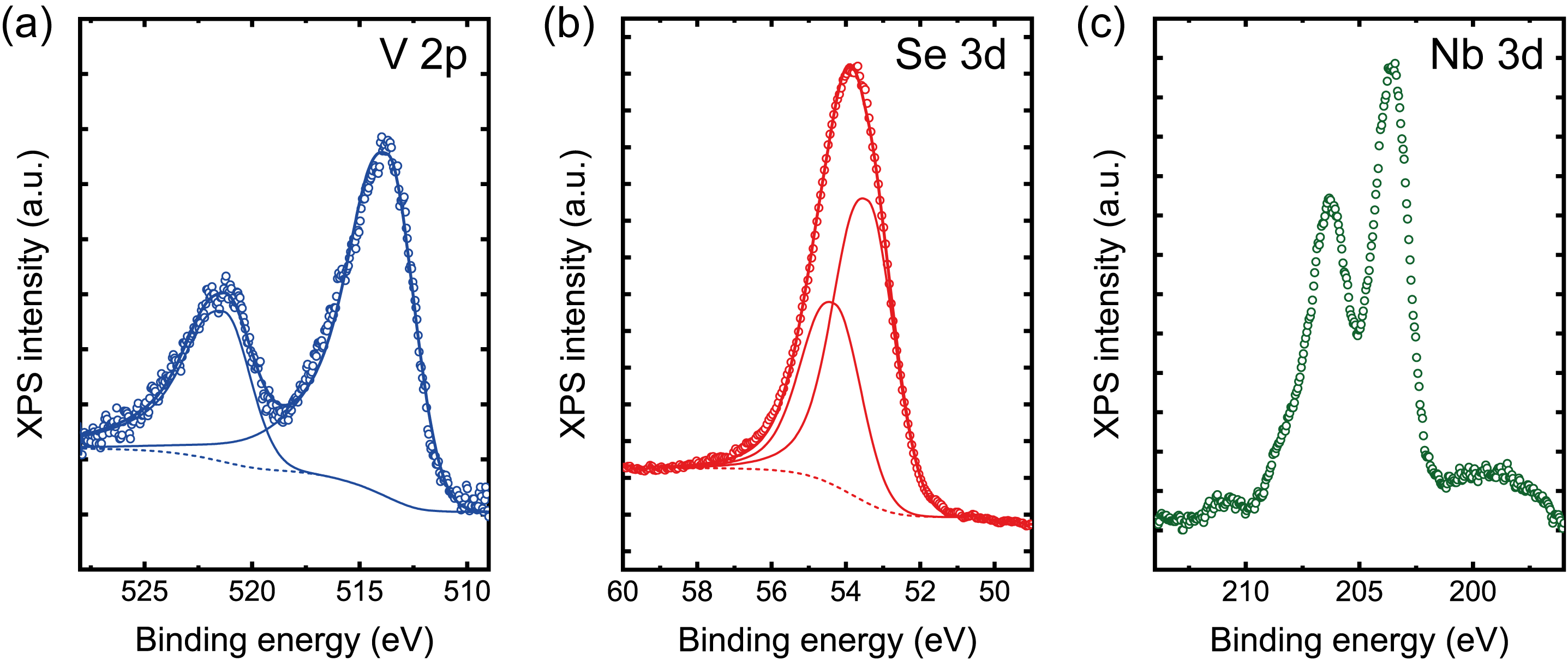}
	\caption{X-ray photoelectron spectra from a 0.6 ML as-grown VSe$_2$ film on NbSe$_2$. (a) V 2p region. (b) Se 3d region fitted with two Gaussian/Lorentzian peaks corresponding to the 3d$_{5/2}$ and 3d$_{3/2}$ peaks. (c) Nb 3d region.}
	\label{figS1}
\end{figure}

\section{Computational details}

All density-functional theory calculations are carried out 
in the plane-wave basis in the projector augmented wave framework
as implemented in VASP \cite{kres1,kres2,PAW}. In all calculations, we use 500 eV cutoff and
$k$-point sampling corresponding to $24\times24$ mesh in the primitive cell.
High $k$-point mesh is required to correctly describe e.g. the CDW phases.

V tends to exhibit strong Coulomb correlations, which usually
necessitates using either hybrid functionals or +U. 
It was shown in Ref.\ \cite{Fumega18},
the depending on the U-parameter, the monolayer is either
in nonmagnetic CDW state or in ferromagnetic no-CDW state.
This balance is also affected by the number of electrons \cite{Fumega18}
and likely also by strain.

\begin{figure}[!ht]
\begin{center}
  \includegraphics[width=14cm]{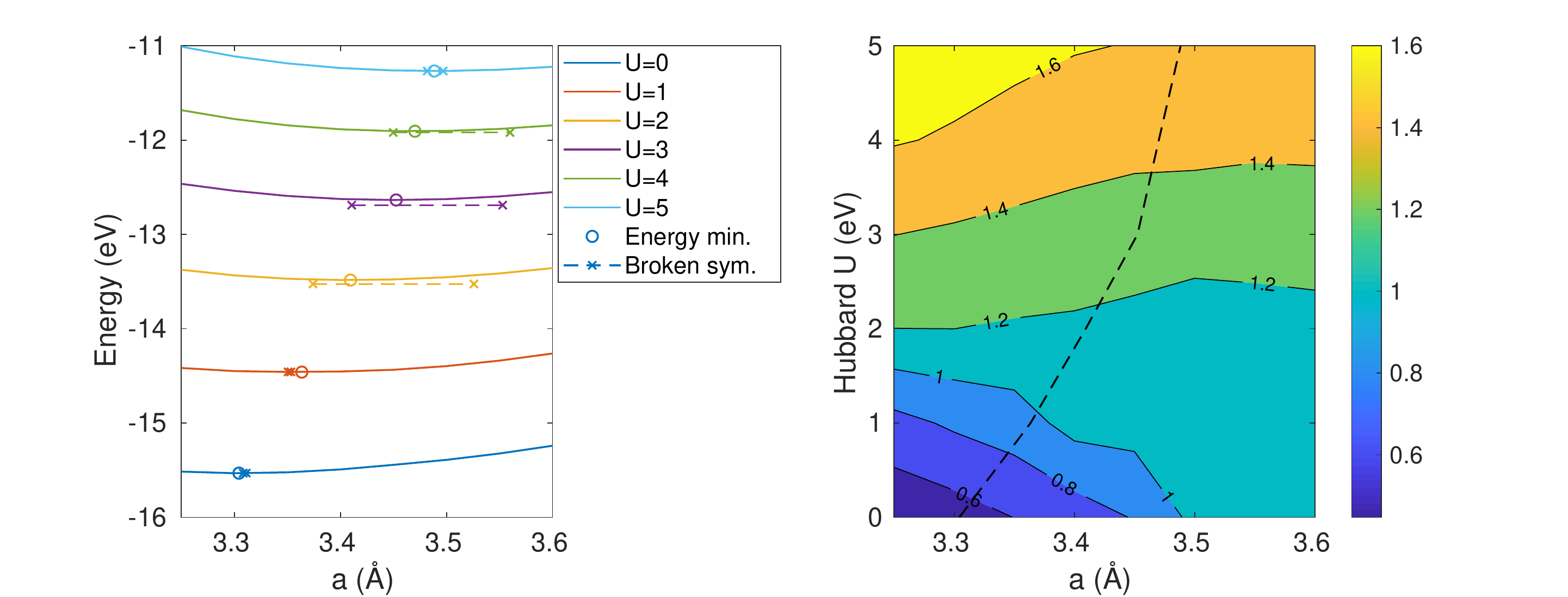}
\end{center}
\caption{\label{fig:uascan}
a) Total energy vs. lattice constant for different values of U-parameter
calculated using revB86b.
The energy minimum position is highlighted with a circle. The lattice constants
and energy of the broken symmetry configuration are shown with crosses.
b) Map of magnetic moments in the U/$a$ space. 
Dashed line denotes the optimal lattice constants.
}
\end{figure}

We calculated the lattice constants and magnetic moments using several exchange-cor\-re\-la\-tion
functionals  \cite{Perdew96,Grimme06,Hamada14_PRB} in order to choose a suitable one.
The results are shown in Table \ref{tab:latt} and also in Fig.\ \ref{fig:uascan}.
In the case of H-NbSe$_2$, all three tested functionals appear to give reasonable
description of the lattice constant. 
In the case of T-VSe$_2$, we first note, that the bulk lattice constant (3.355 {\AA})
is smaller than the one measured here for monolayer from STM images (about $3.5\pm0.1$ {\AA}),
which could be due to the transition between FM/CDW phases.

LDA strongly underestimates all lattice constants, but 
PBE, PBE-D2, and revB86b all slightly underestimates the lattice constant, although usually 
expected to overestimate. In all cases, adding Hubbard-like on-site Coulomb interaction
remedies the situation. At moderate values of the U-parameter (1-2 eV),
the lattice constants are close to the HSE value and between the two experimental values,
and also the magnetic moments are close to the HSE value.
At large U, both the lattice constant and the magnetic moment increase.

\begin{table}[h]
\centering
\caption{
The calculated lattice constants and magnetic moments of monolayer NbSe$_2$ and VSe$_2$. 
The experimental lattice constants are also shown and taken from the bulk. 
$\Delta E_{\rm CDW}$ is the energy difference between $\sqrt{3}$R30$\times \sqrt{7}$R19.1
CDW phase and FM phase (positive value means FM phase is lower in energy).
}
\label{tab:latt}
\begin{tabular}{lccccc}
\hline 
		& \multicolumn{2}{c}{H-NbSe$_2$}	& \multicolumn{3}{c}{T-VSe$_2$} \\
		& a	& M	& a	& M & $\Delta E_{\rm CDW}$ (eV) \\
\hline
Expt. a/c	& 3.442	&	& 3.355	& & \\
\hline
PBE 		& 3.486	&	& 3.336	& 0.6 & 0.013 \\
PBE(U=1)	&	&	& 3.405 & 1.1 & \\
PBE(U=2)	&	&	& 3.406	& 1.2 & \\
PBE-D2		& 3.467	&	& 3.321	& 0.6 & \\
revB86b		& 3.461	&	& 3.306	& 0.5 & 0.004 \\
revB86b(U=1)	&	&	& 3.356	& 0.9 & \\
revB86b(U=2)	&	&	& 3.408	& 1.2 & 0.161 \\
revB86b(U=3)	&	&	& 3.451	& 1.3 & \\
revB86b(U=4)	&	&	& 3.468	& 1.5 & \\
revB86b(U=5)	&	&	& 3.488	& 1.6 & \\
LDA		& 	&	& 3.220 & 0.2 & -0.007 \\
HSE		&	&	& 3.396	& 1.0 & \\
\end{tabular} 
\end{table}

\begin{figure}[!ht]
\begin{center}
  \includegraphics[width=18cm]{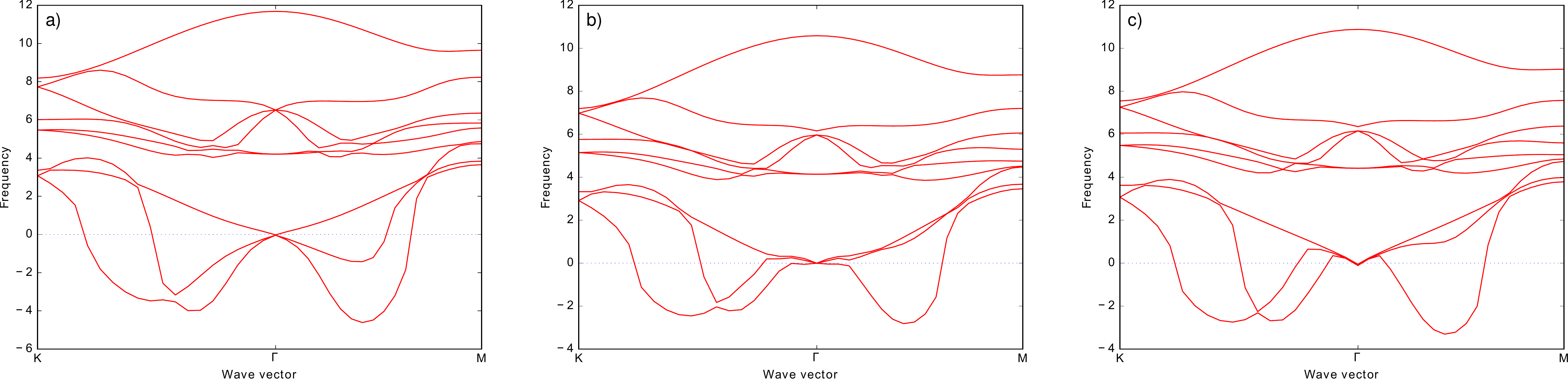}
\end{center}
\caption{\label{fig:phonon}
Phonon dispersion spectra in the NM phase calculated using
a) LDA, b) PBE, and c) revB86b.
}
\end{figure}

We also studied the CDW phase stability. We here only considered
the $\sqrt{3}$R30$\times\sqrt{7}$R19.1 pattern suggested in Ref.\ \cite{Coelho19_JPCC}.
As shown in the calculated phonon dispersion curves in Fig.\ \ref{fig:phonon},
the imaginary frequency modes are located at the same wavevectors, which
justifies using the same CDW pattern with all functionals.
The energy difference between this phase and the FM phase are also given
in Table \ref{tab:latt}.
Only within LDA, CDW phase is slightly lower in energy (by 7 meV/formula unit),
whereas PBE and revB86b both slightly favor FM phase (by 4--13 meV/f.u.).
Upon adding +U correction the FM phase becomes strongly favored.

We note, that for U values in the range 2--4 eV, a lower energy FM structure with
broken symmetry was found (cf.\ Fig.\ \ref{fig:uascan}). 
The unit cell size remains at 3 atoms, but the two lattice constants have
different lengths.

As a result, revB86b with moderate U (e.g. 2 eV) gives 
results in good agreement with HSE and also lattice constant in fairly good
agreement with experiments.
On the other hand, the CDW phase can be stabilized, or nearly
stabilized, only without U.


\subsection{Electronic structure}

\begin{figure}[!ht]
\begin{center}
  \includegraphics[width=16cm]{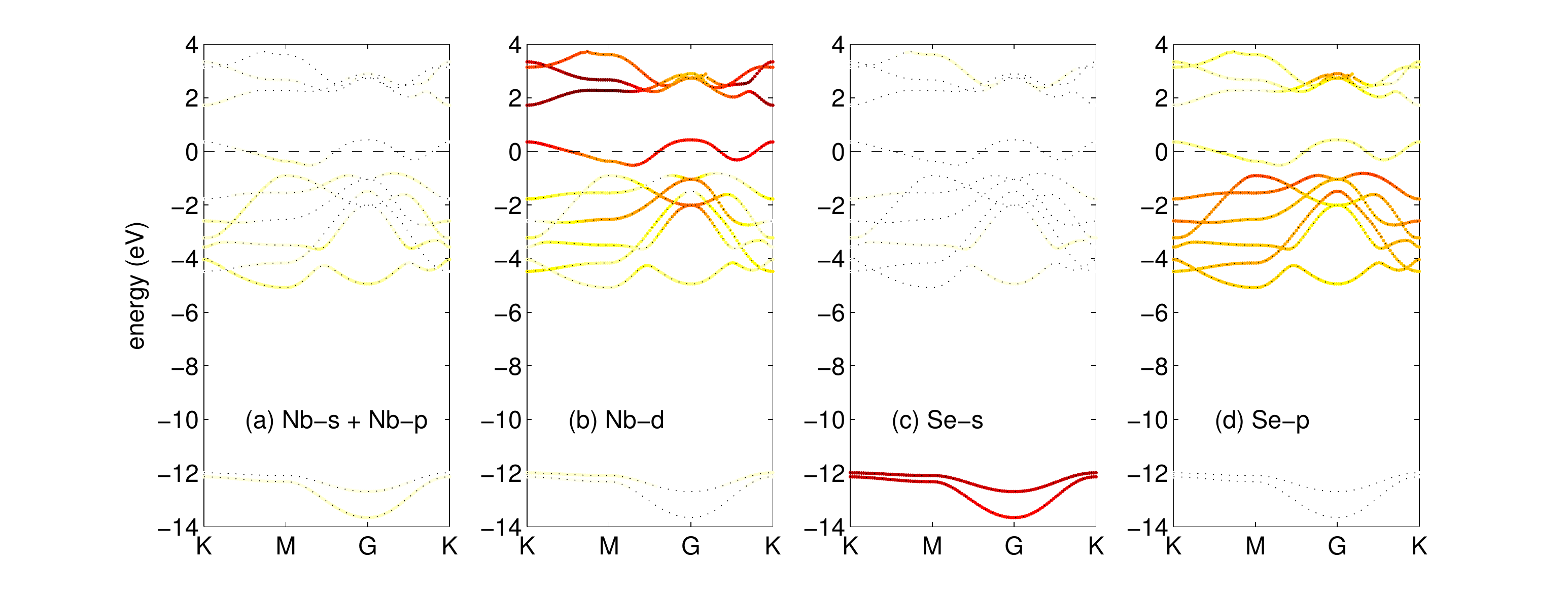}
\end{center}
\caption{\label{fig:bsNb}
Band structure of monolayer H-NbSe$_2$ projected
on (a) Nb-sp, (b) Nb-d, (c) Se-s, and (d) Se-p.
Energy zero is at Fermi-level.
}
\end{figure}

Band structure of monolayer NbSe$_2$ is shown in Fig.\ \ref{fig:bsNb}.
The valence bands are all formed out of Nb-d and Se-p.
Fermi-level crosses one distinct band, which is also responsible
for the CDW. 
Spin-orbit coupling splits this band by 0.19--0.31 eV.

\begin{figure}[!ht]
\begin{center}
  \includegraphics[width=16cm]{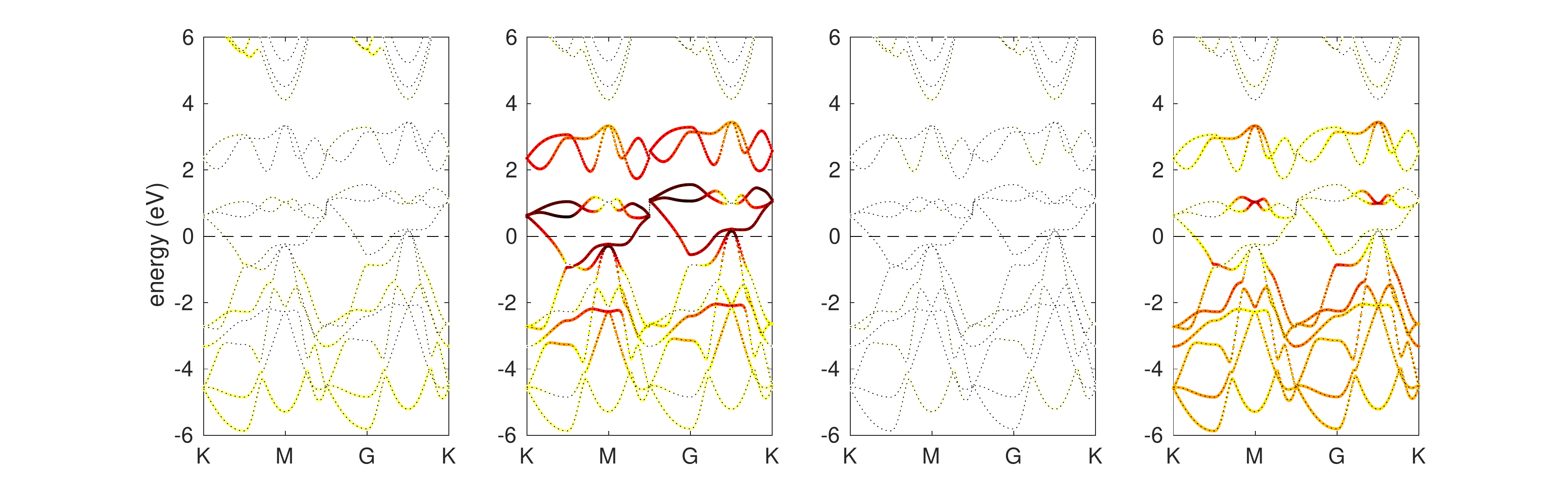}
  \includegraphics[width=16cm]{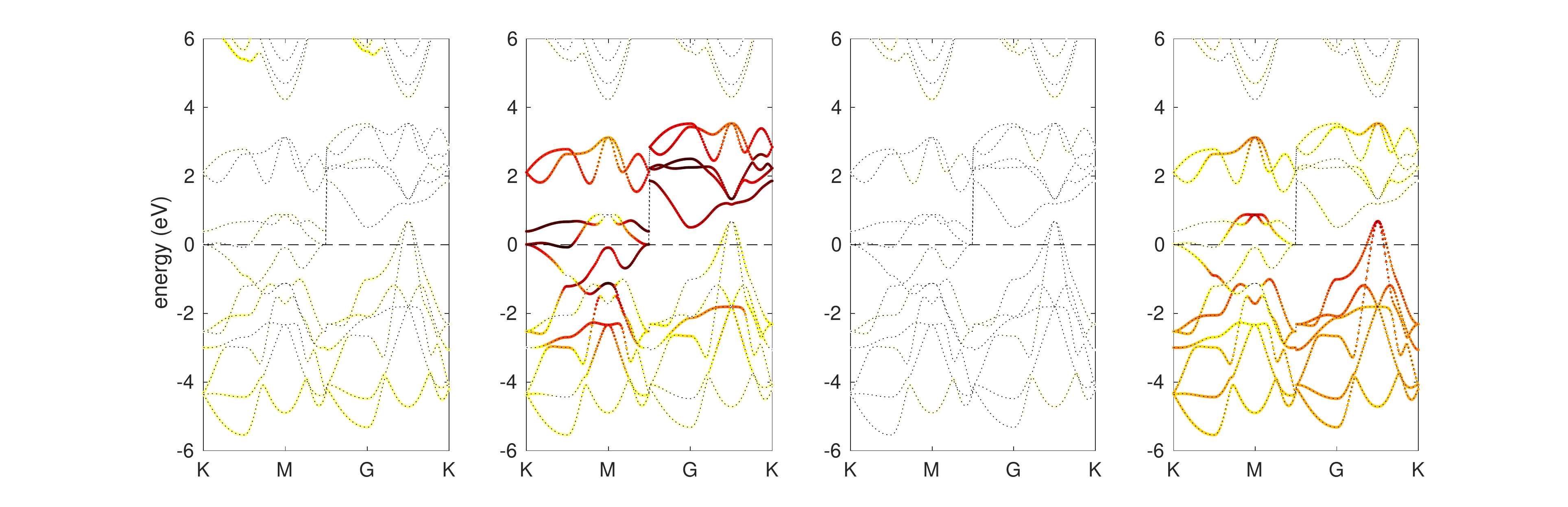}
  \includegraphics[width=16cm]{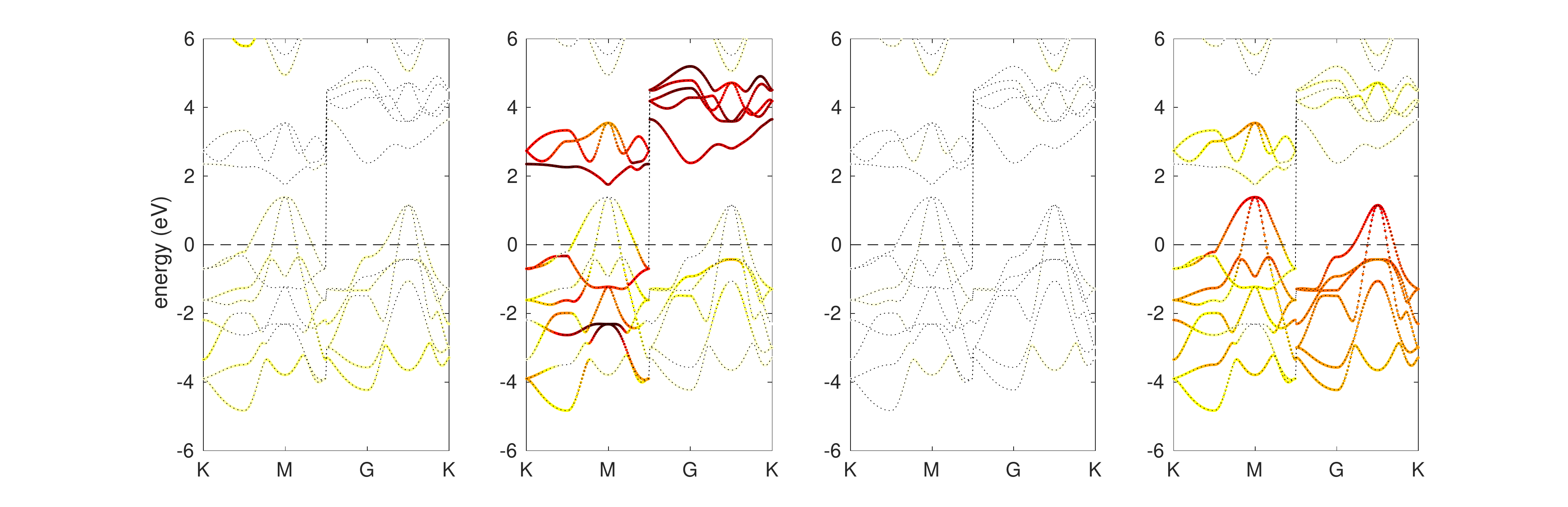}
\end{center}
\caption{\label{fig:bsV}
Band structure of monolayer T-VSe$_2$ calculated using revB86b (top), revB86b+U(2) (middle), and  revB86b+U(5) (bottom), 
projected on V-sp, V-d, Se-s, and Se-p, respectively in
the four columns.
Half of k-points are spin up and half are spin down.
Energy zero is at Fermi-level.
}
\end{figure}

Band structure of monolayer VSe$_2$ is shown in Fig.\ \ref{fig:bsV}.
There is partly filled V-d band, similar to NbSe$_2$.
+U parameter controls the separation between the spin-up and -down channels of this
band.

The PBE calculated work functions are 5.56 eV for NbSe$_2$
and 5.02 eV for VSe$_2$.
Thus, we expect to have electron transfer from VSe$_2$ to NbSe$_2$
(hole-doping of VSe$_2$).
Comparing these to the Fermi-level position of graphene at 4.39 eV suggests charge transfer from graphene to VSe$_2$
(electron-doping of VSe$_2$).
According to Fumega and Pardo \cite{Fumega18}, FM becomes stronger at small hole-doping
and weaker at electron-doping (assuming Stoner mechanism).

\subsection{CDW structures}

Bulk NbSe$_2$ adopts incommensurate CDW phase,
although nearly commensurate to 3$\times$3 \cite{Malliakas13_JACS}, 
at $T<33$ K.
Calculations are carried out at 0 K and thus also show
CDW-phase when suitable supercell is used.
In 3$\times$3 monolayer supercell, 
PBE and revB86b calculations yield CDW-phase lower in energy by 4 meV (33 K$\cdot k_B$) per
formula unit. 

\begin{figure}[!ht]
\begin{center}
  \includegraphics[width=9.5cm]{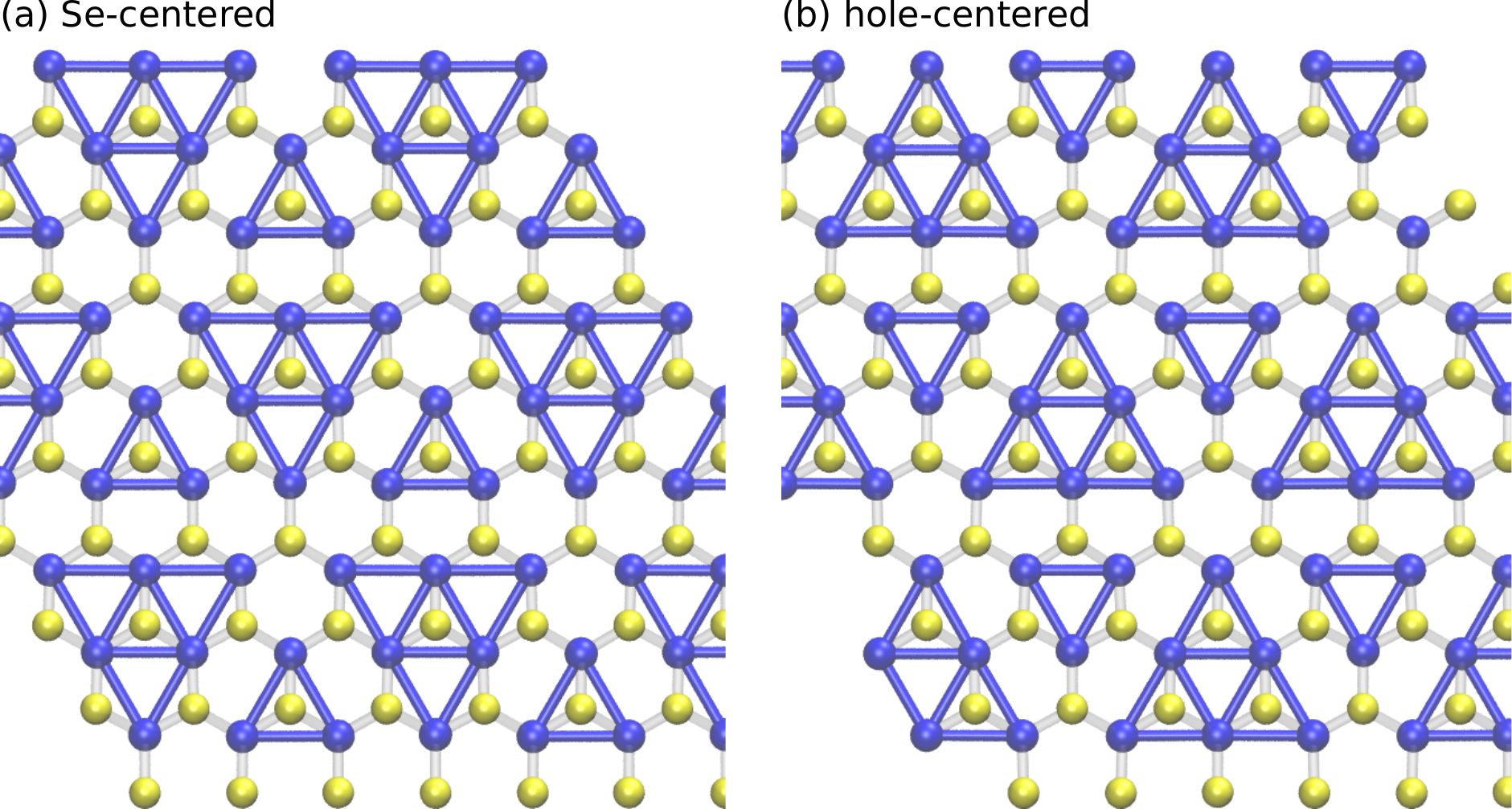}
\end{center}
\caption{\label{fig:cdw}
Atomic structure of NbSe$_2$ in CDW phase.
}
\end{figure}

There are two nearly degenerate phases for the CDW distortion,
hexagon-centered (HC) and anion-centered (AC) \cite{Gye19_PRL}.
These are shown in Fig.\ \ref{fig:cdw}.

\begin{figure}[!ht]
\begin{center}
  \includegraphics[width=6cm]{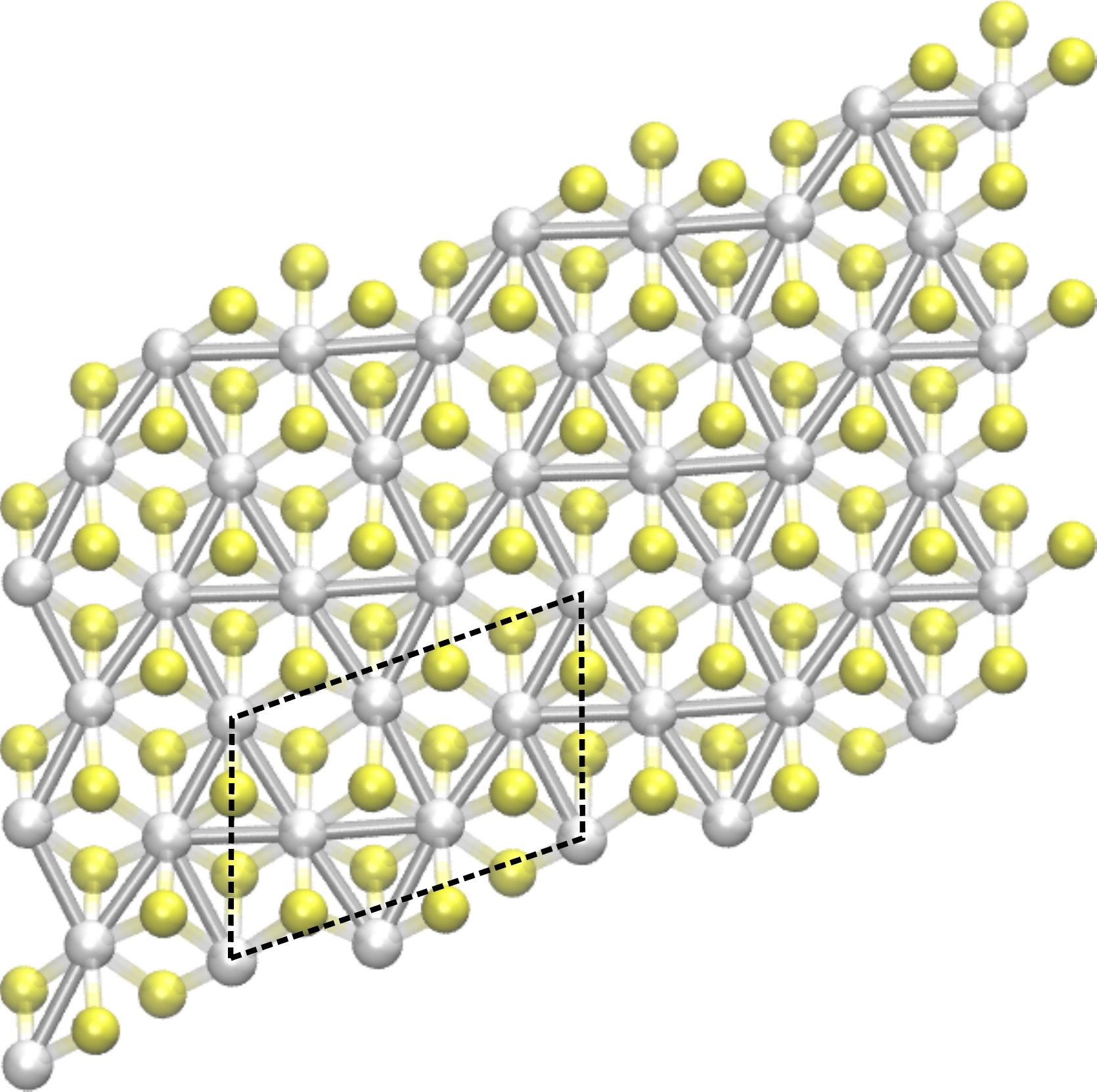}
\end{center}
\caption{\label{fig:cdwvse}
Atomic structure of VSe$_2$ in CDW phase.
Dashed parallelogram depicts the CDW unit cell.
}
\end{figure}

Atomic structure of the CDW phase VSe$_2$ is shown in Fig.\ \ref{fig:cdwvse}.

\subsection{Heterostructure}

In bulk H-NbSe$_2$, Nb atoms are on top of each other and Se atoms positions alternate.
Bulk T-VSe$_2$ adopts an AA stacking.

The experimentally measured lattice constants are very similar,
and thus it should be safe to use just a unit cell when studying
the stacking in the NbSe$_2$/VSe$_2$ heterostructure.

Fixing H-NbSe$_2$ to BaB, T-VSe$_2$ can be located
at BaC, CbA, and AcB, or inverted at CaB, AbC, and BcA.
The binding energies are listed in Table \ref{tab:stack},
and defined with respect to strained monolayers, so that it only contains
the interlayer binding contribution.
The results without +U show large changes in the magnetic moment,
indicating sensitivity of the system to converge to either FM or NM phase.

\begin{table}[h]
\centering
\caption{
Properties of the H-NbSe$_2$/T-VSe$_2$ heterostructures:
interlayer binding energy $E_b$, total magnetization per unit cell $M$, distance between the Nb and V layers $h_{\rm Nb-V}$, 
and the dipole over the heterostructure $p$.
The first set is without +U correction and the second with $U=2$ eV.
}
\label{tab:stack}
\begin{tabular}{ccccccc}
\hline 
	       & AbC	& AcB	& BaC	& BcA	& CaB	& CbA \\ 
\hline 
$E_b$ (eV)	&-0.249	&-0.272	&-0.175	&-0.144	&-0.287	&-0.222 \\
$M$ ($\mu_B$)	& 0.96	& 0.92	& 0.95	& 0.0	& 0.95	& 0.0 \\
\hline
$E_b$ (eV)	&-0.239	&-0.259	&-0.166	&-0.168	&-0.269	&-0.250 \\
$h_{\rm Nb-V}$	& 6.34	& 6.23	& 6.76	& 6.76	& 6.16	& 6.24 \\
$p$ (meV{\AA})	& 14.0	& 14.2	& 14.5	& 14.1	& 14.4	& 14.6 \\
$M$ ($\mu_B$)	& 1.19	& 1.14	& 1.15	& 1.16	& 1.14	& 1.20 \\
\hline
\hline
\end{tabular} 
\end{table}

The preferential stacking has Se of VSe$_2$ on top of hexagon hole of NbSe$_2$
and V on top of Nb (CaB).
The binding is clearly weakest when the Se of VSe$_2$ is on top of Se of NbSe$_2$.
The binding energies clearly correlate with the interlayer distance.
Strongest binding energy corresponds to closest distance.

The charge transfer and dipole over the heterostructure seem similar in all
cases, and generally quite small, e.g., 0.01 e from VSe$_2$ to NbSe$_2$ in
CaB case (about $10^{13}$ cm$^{-2}$).
Such charge transfer should be insufficient to destroy CDW ordering of the NbSe$_2$
under VSe$_2$.

\begin{figure}[!ht]
\begin{center}
  \includegraphics[width=16cm]{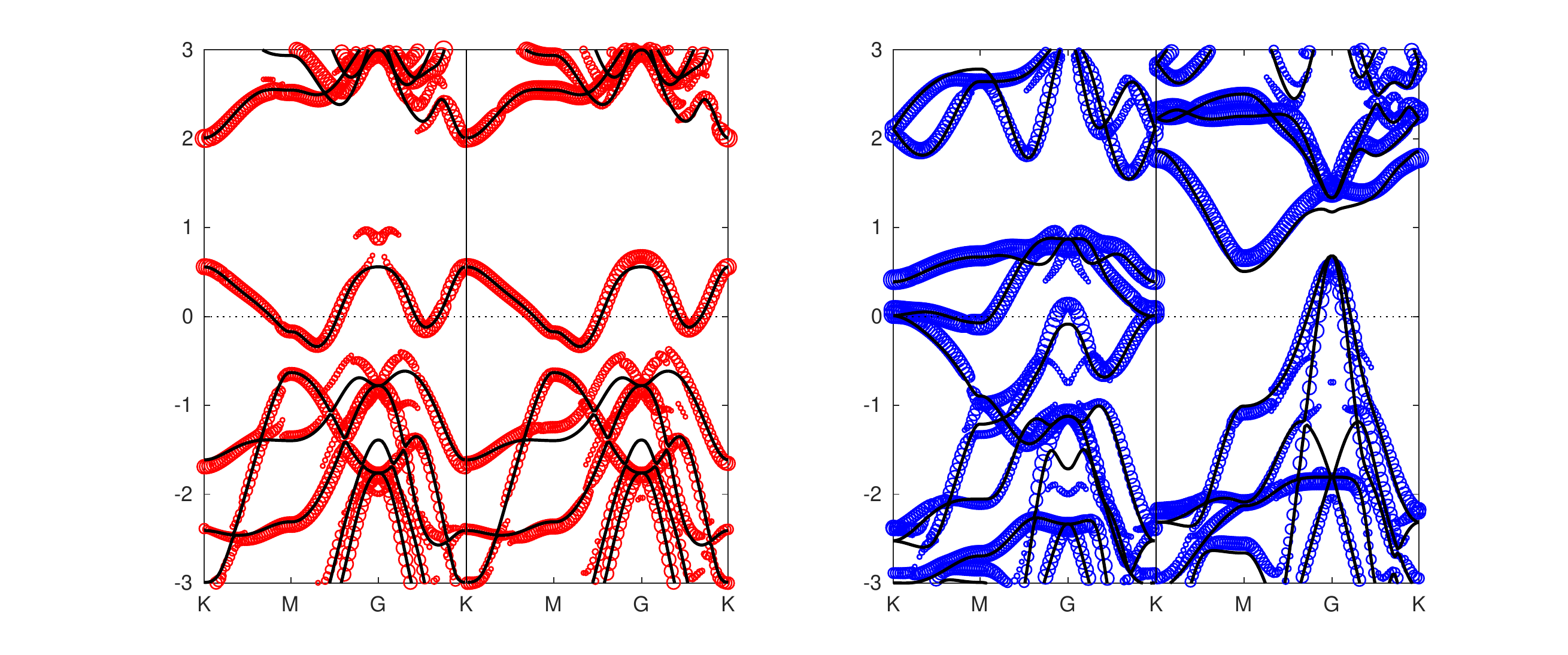}
\end{center}
\caption{\label{fig:bshs}
Band structure projected to NbSe$_2$ (left, red) and VSe$_2$ (right, blue) layers
from the BaB/CaB stacked heterostructure.
The band structure from pristine monolayers are shown with black lines.
Fermi-level is set to zero in all band structures.
Left/right half is spin-up/down.
}
\end{figure}

Fig.\ \ref{fig:bshs} shows the BaB/CaB heterostructure 
band structure projected to NbSe$_2$ and VSe$_2$ layers,
calculated using revB86b+U(2).
There are no dramatic changes due to stacking.
By comparing to the orbital projections in Figs.\ \ref{fig:bsNb} and \ref{fig:bsV}
it can be seen that the largest changes occur whenever the bands have
strong Se-p contribution, which can feel the other layer.

\begin{figure}[!ht]
\begin{center}
  \includegraphics[width=16cm]{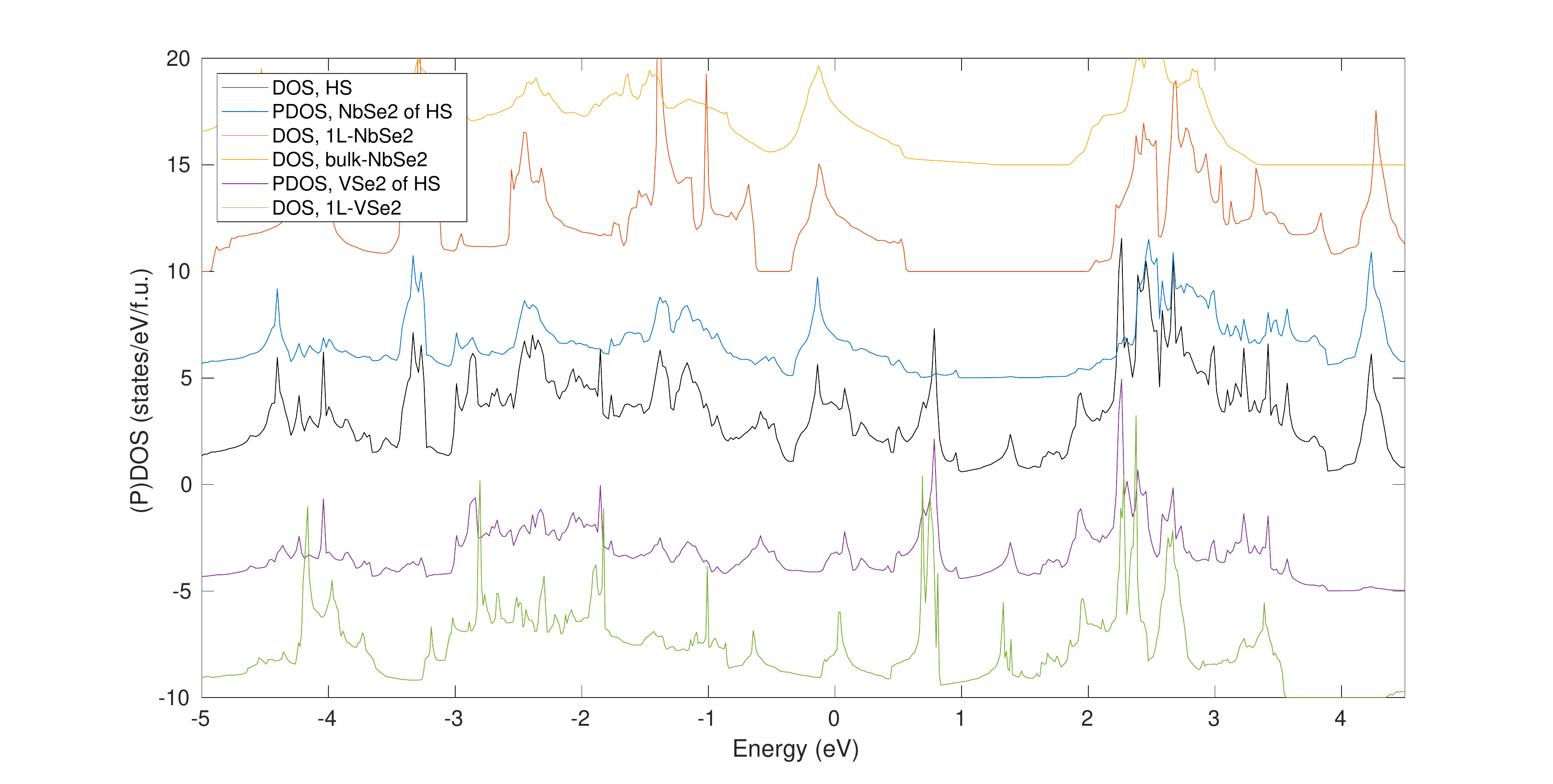}
\end{center}
\caption{\label{fig:doshs}
DOS of the heterostructure is shown in the middle. Above/below are the contributions to it from NbSe$_2$/VSe$_2$ layers, and above/below those are the DOSes of
pristine NbSe$_2$/VSe$_2$ monolayers for comparison.
The spin-up and -down channels are summed up.
Fermi-level is set to zero.
}
\end{figure}

Densities of states (DOS) of the heterostructure and
the isolated monolayers are shown in Fig.\ \ref{fig:doshs}.

\subsection{Simulated STM images}

\begin{figure}[!ht]
\begin{center}
  \includegraphics[width=16cm]{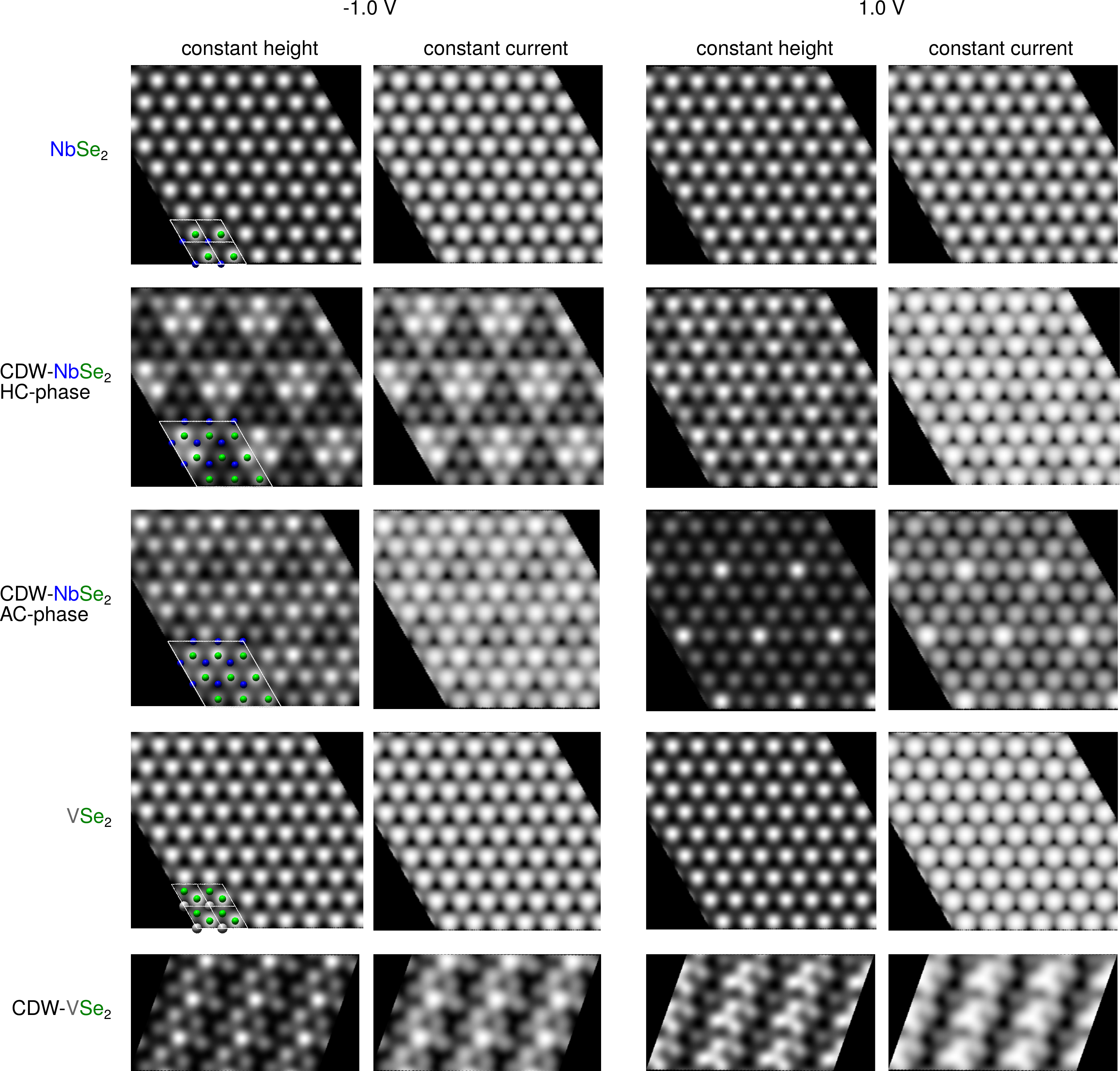}
\end{center}
\caption{\label{fig:stm}
STM images of (a) pristine no-CDW NbSe$_2$, (b) CDW-NbSe$_2$ hexagon-centered phase,
(c) CDW-NbSe$_2$ anion-centered phase, (d) no-CDW VSe$_2$, and (e) CDW-VSe$_2$.
The atomic structures are overlaid.
}
\end{figure}

STM images of the NbSe$_2$ and VSe$_2$ are shown in Fig.\ \ref{fig:stm}.
The constant height image is recorded at 3 {\AA} from the topmost Se atom
and the constant current image at different small isosurface values,
which is the same for different biases, but different in different materials.
For the non-CDW phases they all look similar, with protrusions at Se atoms.
For the CDW-NbSe$_2$, the experimental images shown in the main text 
have close match with the simulated images in the HC-phase.

The simulated STM image for CDW-VSe$_2$ is in good agreement with the
experimental images reported in \cite{Coelho19_JPCC}, but not observed
in our experiments.

\subsection{Simulated STS}

STS is simulated within Tersoff-Hamann approach using LDOS as
\begin{equation}
I(E) = \sum_{E<E_i<E+\delta E} \int \rho_i(x,y,z=z_0) dx dy
\end{equation}
where $\rho_i=|\Psi_i|^2$ is the partial charge density of $i$th Kohn-Sham orbital,
$z_0$ determines the distance above the surface, and $\delta E$ sets the energy resolution; 
here 20 meV.

\begin{figure}[!ht]
\begin{center}
  \includegraphics[width=16cm]{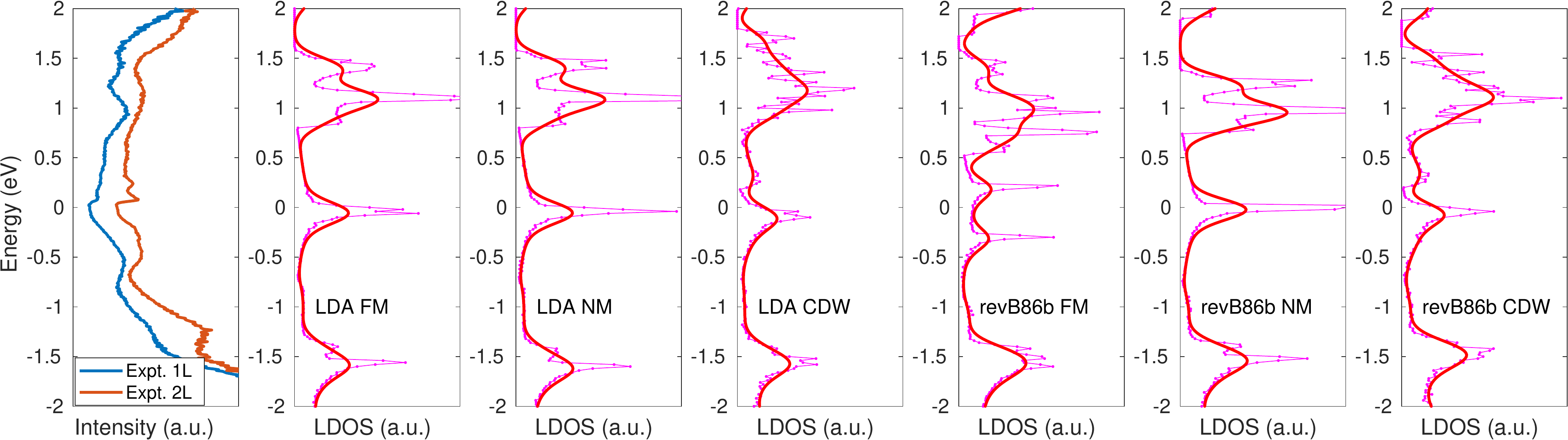}
\end{center}
\caption{\label{fig:stsU}
Comparison of STS spectra evaluated using two different XC functionals and in 
three different phases: ferromagnetic (FM), nonmagnetic (NM), and charge density wave (CDW).
Magenta curves show the bare spectra and red curves have been smoothed to allow
for more direct comparison with the experimental one.
}
\end{figure}

\begin{figure}[!ht]
\begin{center}
  \includegraphics[width=16cm]{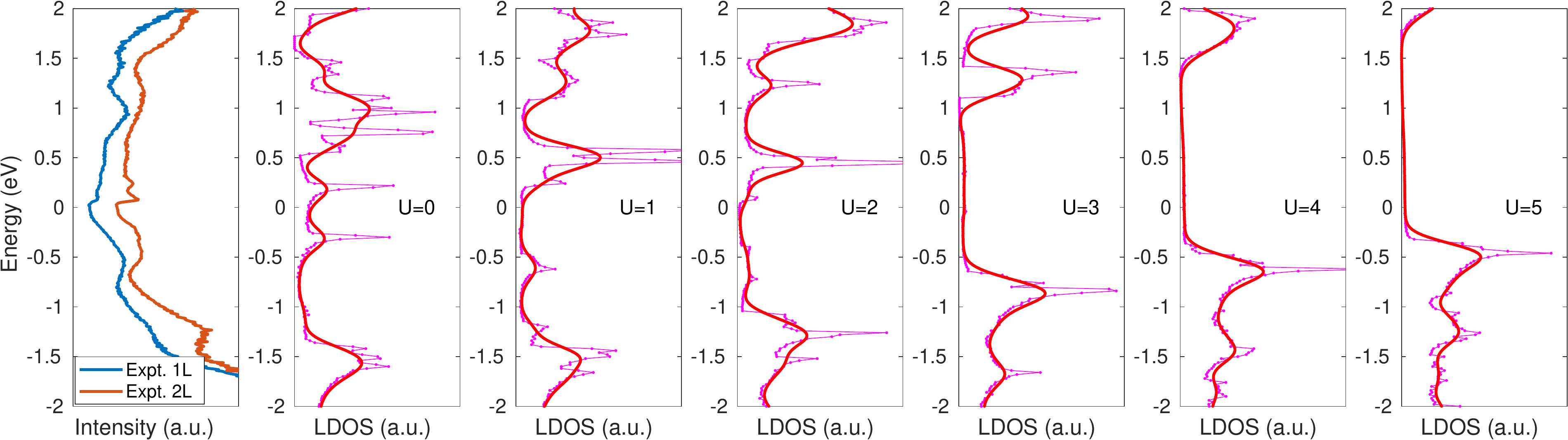}
\end{center}
\caption{\label{fig:stsU}
The evolution of simulated STS spectra with increasing +U parameter (revB86b functional)
and comparison to the experimental spectrum.
}
\end{figure}

LDA and revB86b in the NM phase 
show a very prominent peak at Fermi-level that is in poor agreement with experiments. 
The same is true for the LDA result in FM phase due to very small magnetic moment.
The simulated spectra for the CDW phase seems consistent
with the dip at the Fermi-level and the step-like structure
at positive energies seen in the experimental spectra,
although (i) in the calculations the dip is about 0.15 eV above Fermi-level and (ii) the dip appears to be present already in the NbSe$_2$ spectrum.
Point (i) would contradict with the expected hole-doping from the calculated work functions.
Although point (ii) suggests NbSe$_2$ origin for the dip, the dip becomes more pronounced in the bilayer spectrum. 
These features could be a signature of the CDW phase 
in the bilayer region, which makes sense as bulk VSe$_2$ is known to exhibit CDW and also the bilayer regions are very small and thus strongly affected by the edges.
On the other hand, these features are fairly weak in the monolayer spectrum
and there is also an additional peak at 1.5 eV and the change in the peak
shape at around -0.5 eV.
These changes are consistent with the change between the revB86b calculated
CDW and FM phases. The peak around 1.5 eV becomes more pronounced and
the peak at -0.05 eV in CDW phase moves down to -0.3 eV in the FM phase.

Increasing the value of the U-parameter to 1--2 eV the lower energy
features appear to be in good agreement with experiment,
but there is an extraneous strong peak at about $E_F+0.5$ eV.
At $U \geq 3$ eV, the agreement becomes rather poor.

\begin{figure}[!ht]
\begin{center}
  \includegraphics[width=16cm]{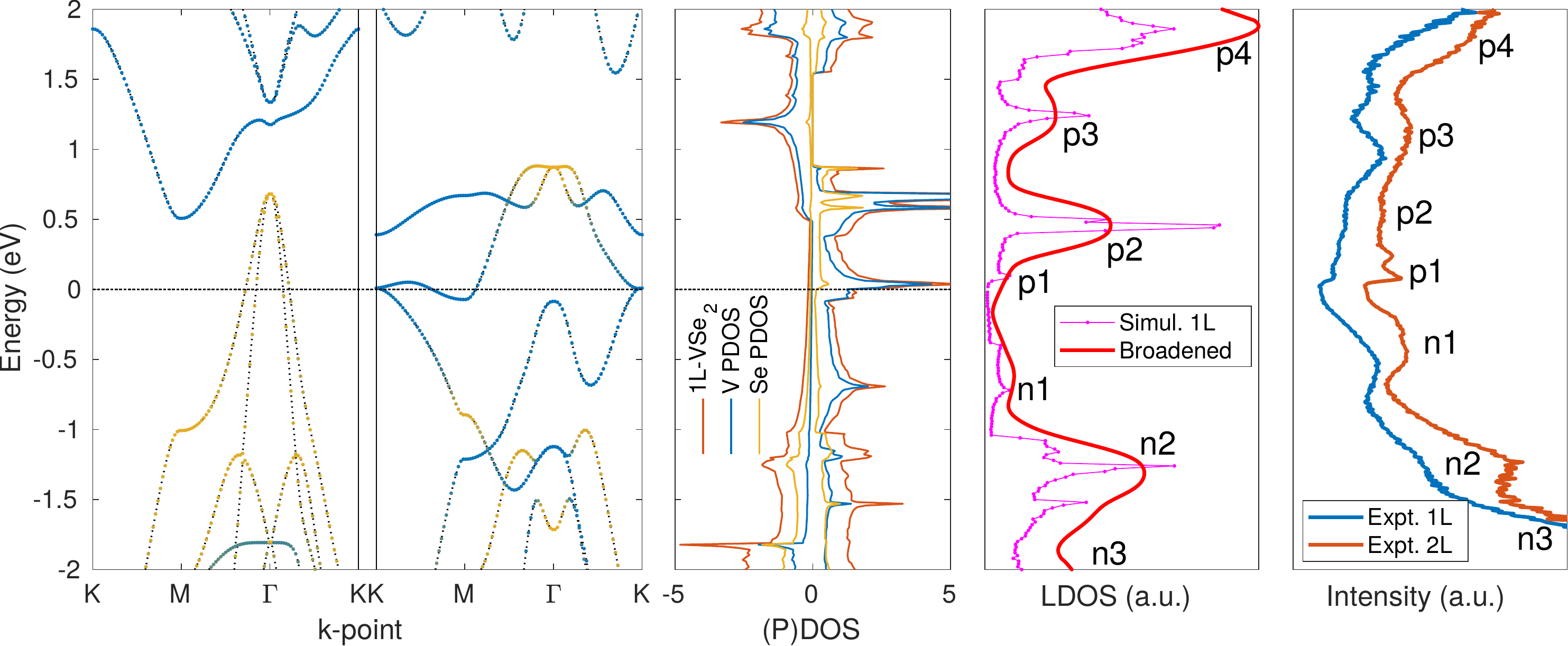}
\end{center}
\caption{\label{fig:dosests}
Band structure, density of states, simulated STS, and experimental STS calculated using revB86b+U(2).
The peaks are denoted as in the main text. 
}
\end{figure}

The band structure, density of states, simulated STS, and experimental STS calculated using revB86b+U(2)
are shown in Fig.\ \ref{fig:dosests}
and can be compared to the revB86b calculated results
in Fig. 3 in the main paper.

\clearpage

\section{Background removal from the VSM measurement}
\begin{figure}[!h]
	\centering
		\includegraphics [width=0.4\textwidth] {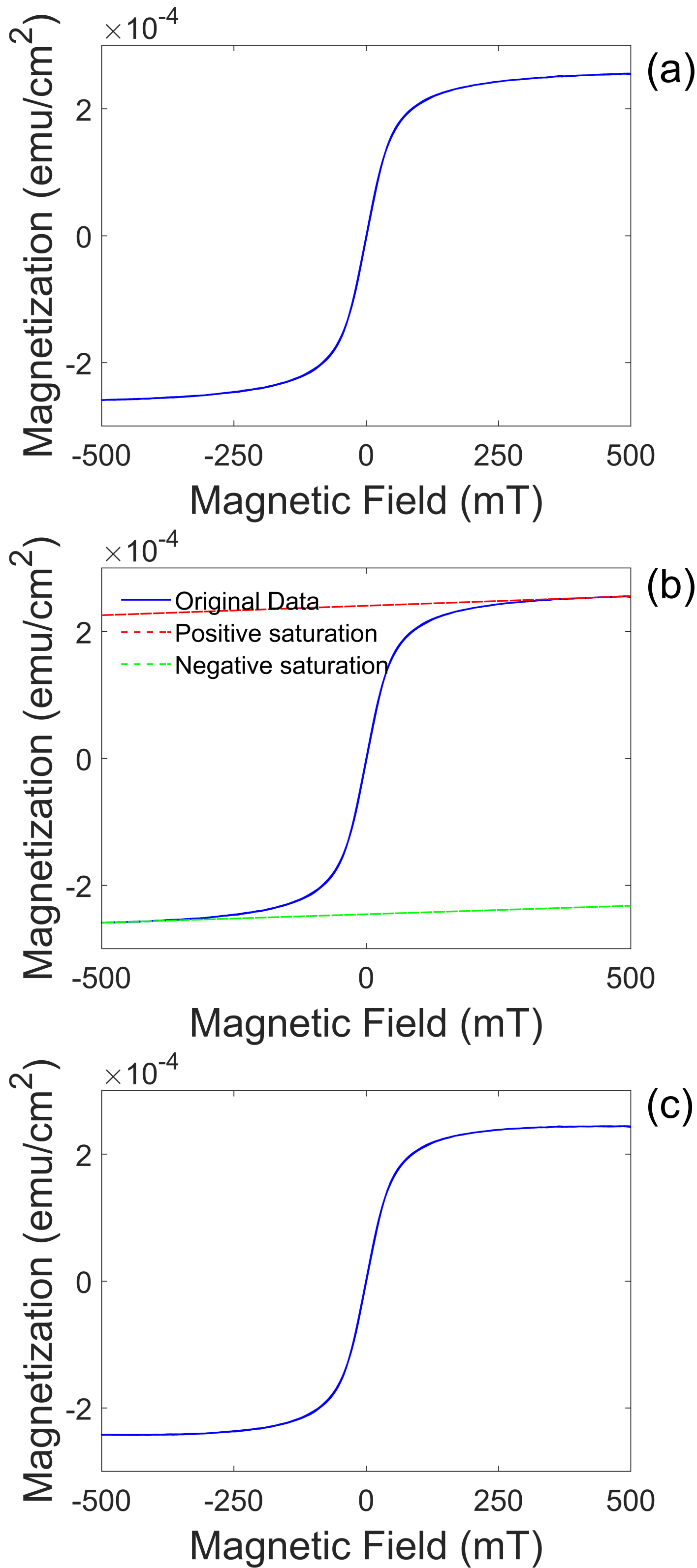}
	\caption{(a) The raw $M-H$ curve of ML VSe$_{2}$ on NbSe$_{2}$ taken at 300 K using vibrating sample magnetometry. (b) The fitted linear background signal and (c) $M-H$ curve with the background subtracted.}
	\label{back}
\end{figure}
 For the data shown in Fig.~3a of the main paper a linear background is removed as shown in Fig.~\ref{back}.

\newpage
\section{Raw data before and after removing monolayer VSe$_2$}
\begin{figure}[!h]
	\centering
		\includegraphics [width=.95\textwidth] {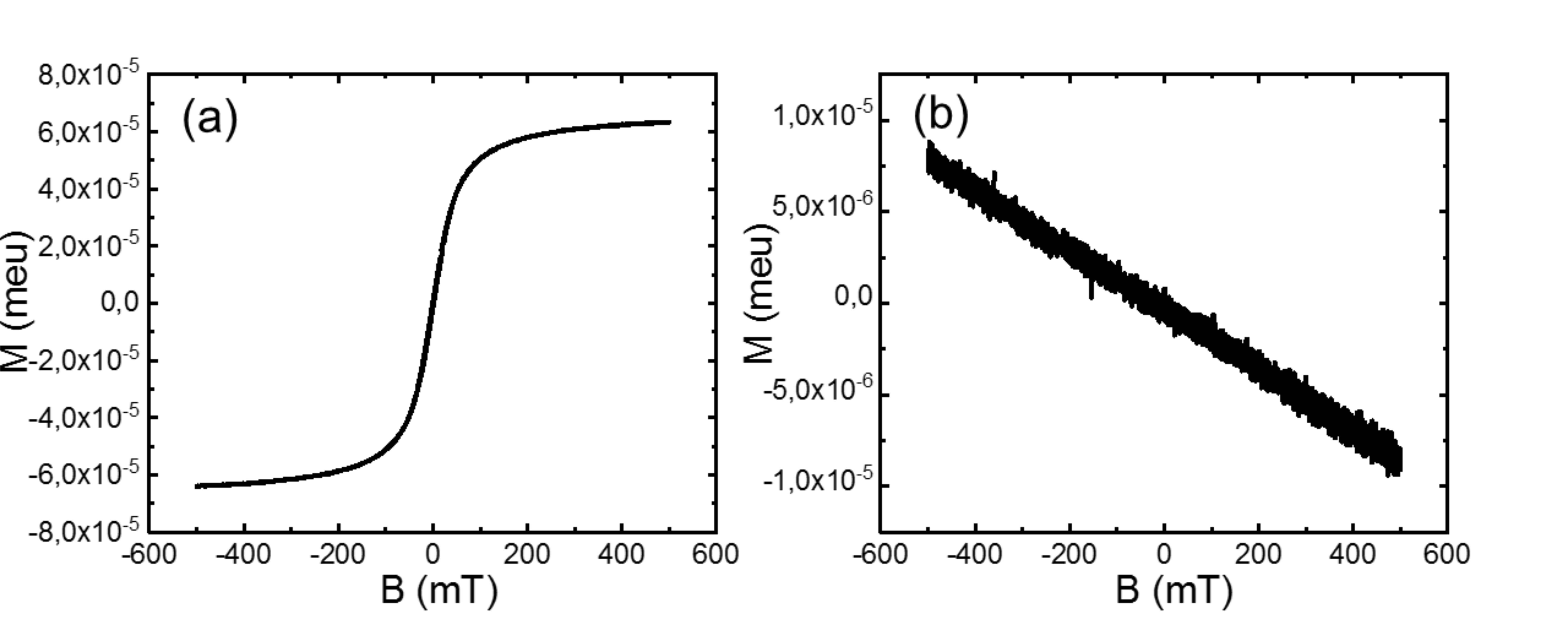}
	\caption{Raw vibrating sample magnetometry (VSM) data before and after removing monolayer VSe$_2$ by cleaving the sample. (a) The raw $M-H$ curve of ML VSe$_{2}$ on NbSe$_{2}$ taken at 300 K using VSM. (b) The raw $M-H$ curve of NbSe$_{2}$ after cleaving the same sample to remove the monolayer VSe$_2$ taken at 300 K using VSM.}
	\label{Figfit}
\end{figure}

\newpage

\section{Temperature dependence of $H-M$ loop}

\begin{figure}[!h]
	\centering
		\includegraphics [width=0.85\textwidth] {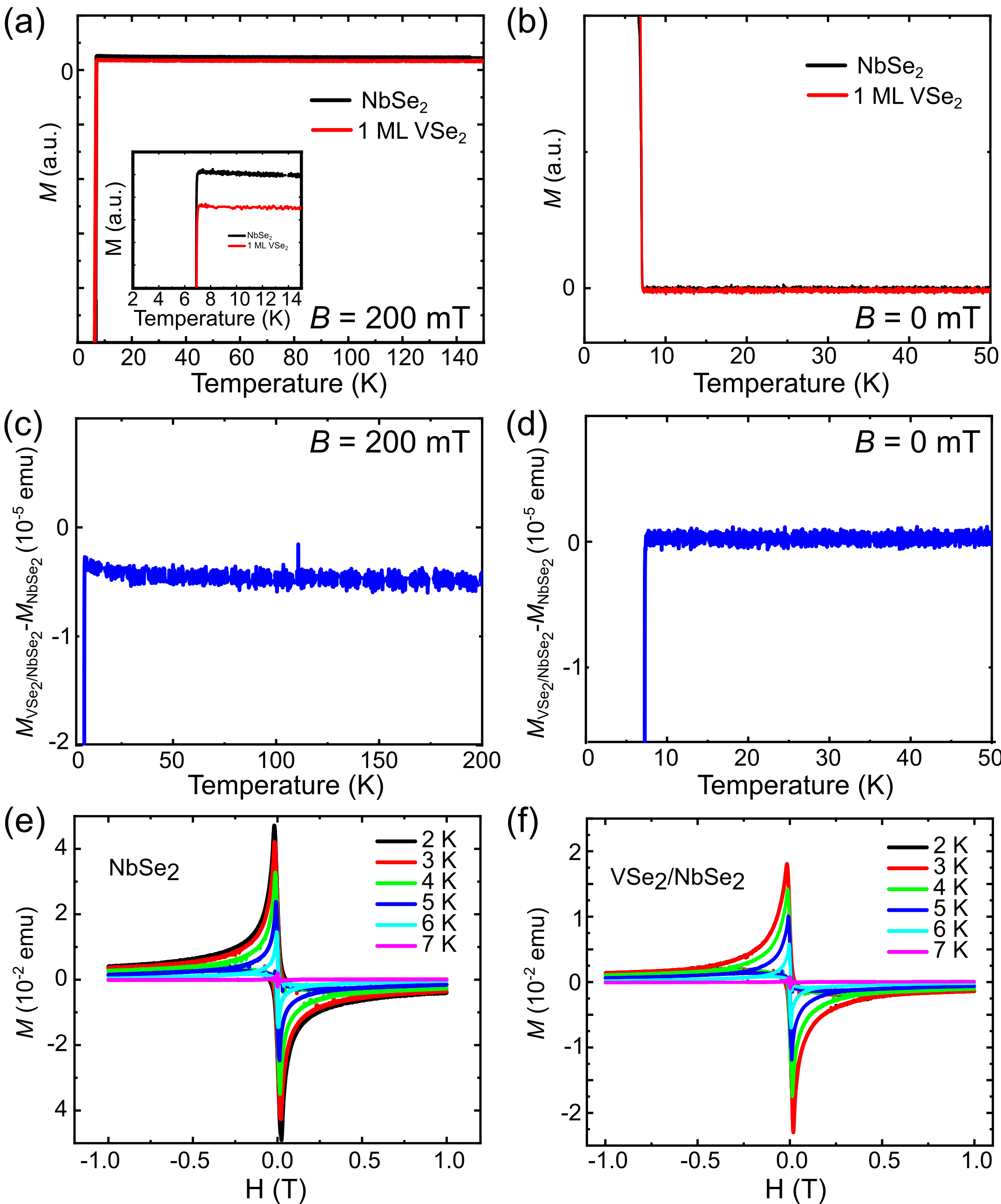}
	\caption{Superconductivity in the VSe$_2$/ NbSe$_2$ hybrids. (a,b) Temperature-dependent field cooling (FC) (magnetic field of 200 mT) and zero-feld cooling (ZFC) magnetization measurements for VSe$_2$ on NbSe$_2$. (c,d)	Background subtracted FC and ZFC (panel c) and zero-field cooling (panel d). (e,f) The temperature dependent $M(H)$ hysteresis loop for the NbSe$_2$ single crystal and VSe$_2$/ NbSe$_2$ hybrids, respectively. }
	\label{FigS4}
\end{figure}

\newpage
\section{Effect of the substrate temperature}
\begin{figure}[!h]
	\centering
		\includegraphics [width=.95\textwidth] {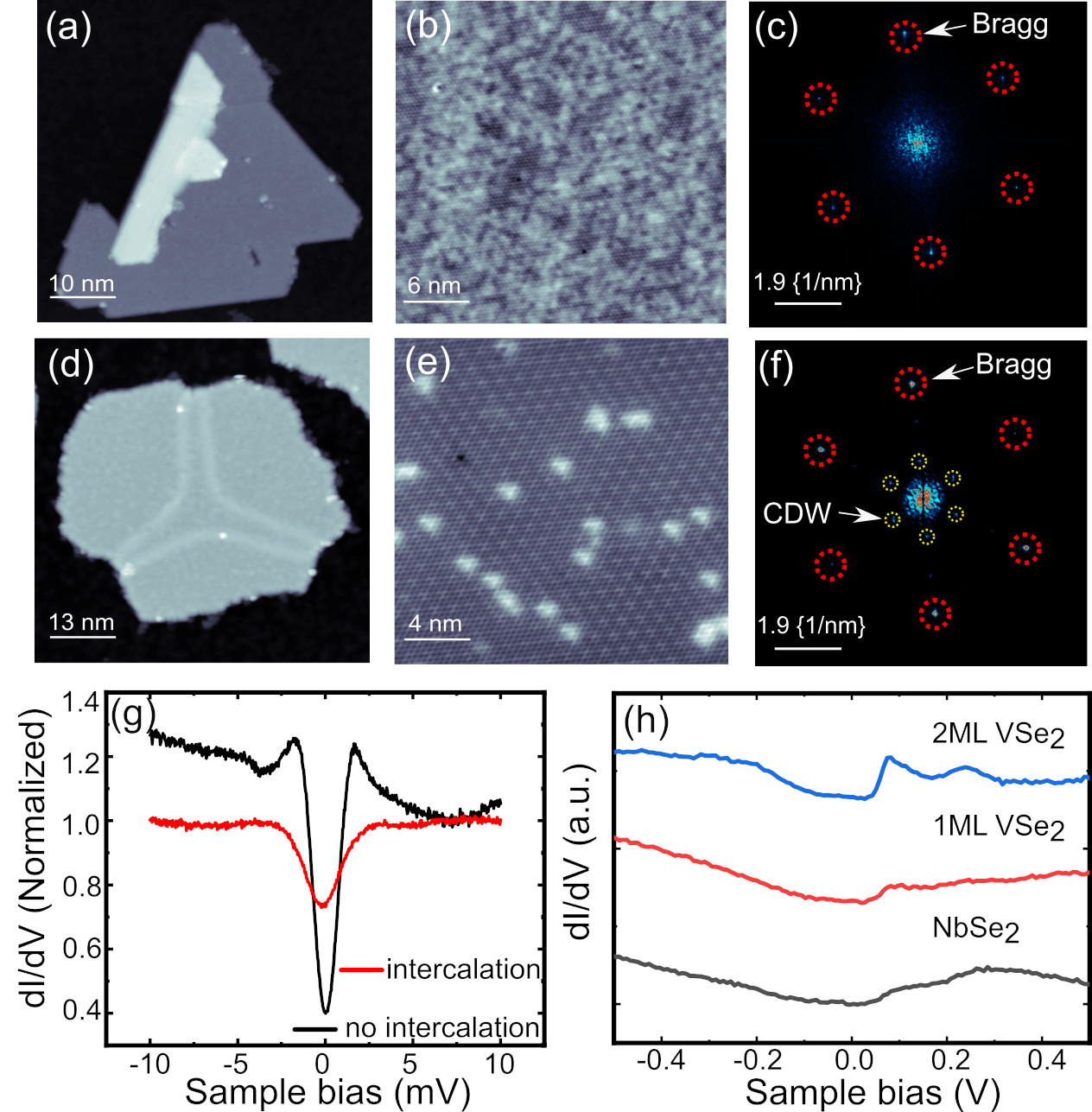}
	\caption{ VSe$_2$ growth on NbSe$_2$ at different substrate temperatures. (a) STM topographic image (1 V, 10 pA) of VSe$_2$ on NbSe$_2$ grown at $T=600$ K. (b) Atomic resolution STM image (10 mV, 1 nA) of NbSe$_2$ surface showing an absence of charge density wave super-structures due to the intercalation. (c) FFT of the atomic resolution image in panel (b). (d) STM topographic image (1 V, 10 pA) of VSe$_2$ on  NbSe$_2$ grown at $T=540$ K. (e) Atomic resolution STM image (10 mV, 1 nA) of NbSe$_2$ surface showing a hexagonal lattic with charge density wave super-structures. (f) FFT of the atomically resolved image shown in panel(e). (f) d$I$/d$V$ spectra measured on the NbSe$_2$ substrate with and without intercalation, respectively. (h) Short-range spectra on the NbSe$_2$ and VSe$_2$ mono- and bilayer showing the absence of a strong CDW gap on VSe$_2$.}
	\label{FigS2}
\end{figure}

The quality of the sample (e.g.~the sharpness of the VSe$_2$ edges) can be improved by growing the sample at a high substrates temperatures. Fig.~\ref{FigS2}a shows STM characterization of sub-monolayer VSe$_2$ films on NbSe$_2$ substrate grown at $T=600$ K. The edge of the island is much sharper compare with low growth temperature, which is grown at $T=540$ K (Fig.~\ref{FigS2}c). However, at high temperature growth, the metal atom starts to intercalate NbSe$_2$ and as a result of that the NbSe$_2$ substrate loses the long-range charge density wave order as shown in an atomic resolution STM image (Fig.~\ref{FigS2}b). In our experiment, we have optimized the substrate temperature ($T=540-570$ K) to insure that the long-range charge density wave order is preserved (Fig.~\ref{FigS2}d). The metal intercalation not only affects the charge density wave order but also the superconducting gap. Fig.~\ref{FigS2}e shows the d$I$/d$V$ spectra measured on NbSe$_2$ surface with and without a metal intercalation. 

\newpage
\section{Fitting the superconducting gap}
While NbSe$_2$ is better described as a two-gap (or anisotropic gap) superconductor, the d$I$/d$V$ spectra at not too low temperatures can be adequately fit with the standard BCS expression  with the density-of-states ($D_s$) given by\cite{hudson1999thesis}
\begin{align}
D_s(\epsilon) = \left\{ \begin{array}{cc} 
                0 & \hspace{5mm} \epsilon<\Delta \\
                \epsilon/(\epsilon^2-\Delta^2)^{1/2} & \hspace{5mm} \epsilon>\Delta \\
                \end{array} \right.
\end{align}
The thermal broadening at temperature $T$ is taken into account in the calculation of the d$I$/d$V$ signal
\begin{equation}
    \frac{\mathrm{d}I}{\mathrm{d}V}(V_b)\propto\int_{\infty}^{\infty}D_s(\epsilon) 
    \frac{\exp((V_b+\epsilon)/kT)}{(1+\exp((V_b+\epsilon)/kT))^2}d\epsilon
\end{equation}
where $V_b$ is the applied bias voltage and $k$ the Boltzmann constant. This expression was then used to fit the experimental d$I$/d$V$ spectra with $\Delta$ as a fitting parameter.
\begin{figure}[!h]
	\centering
		\includegraphics [width=.95\textwidth] {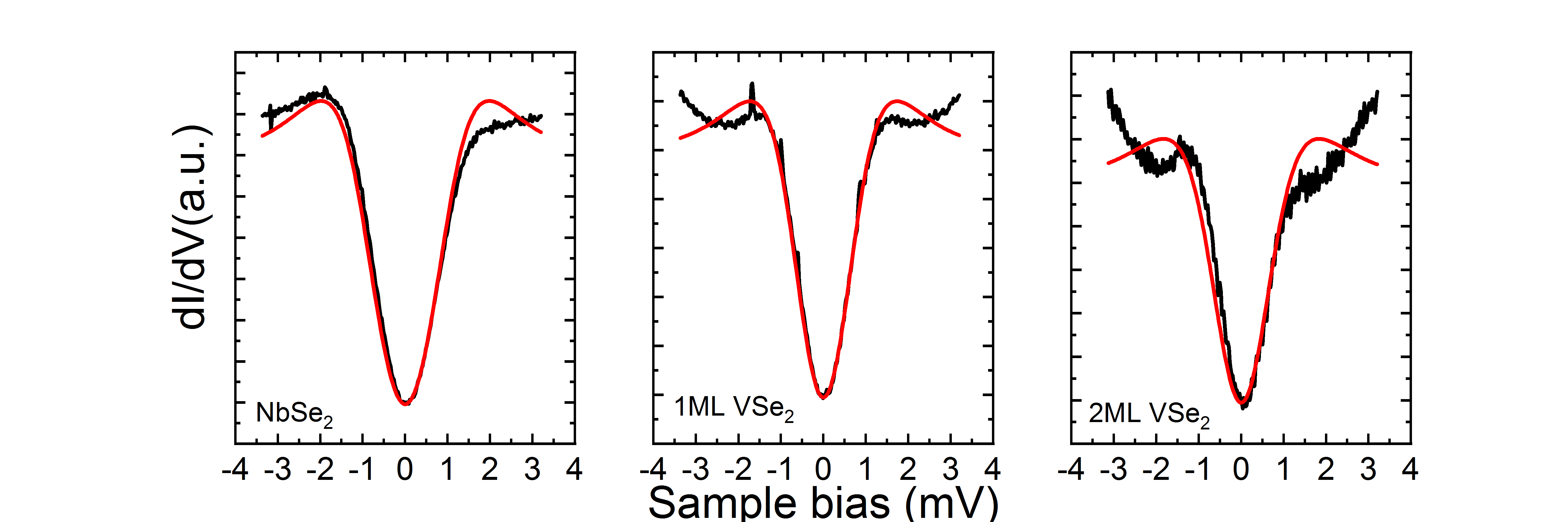}
	\caption{number of layer dependence of the superconducting gap value (black line) obtained from the fit of our STS data to the BCS function (red line)}
	\label{Figfit}
\end{figure}

\newpage
\section{Evolution of the edge state of monolayer VSe$_2$}
\begin{figure}[!h]
	\centering
		\includegraphics [width=0.5\textwidth] {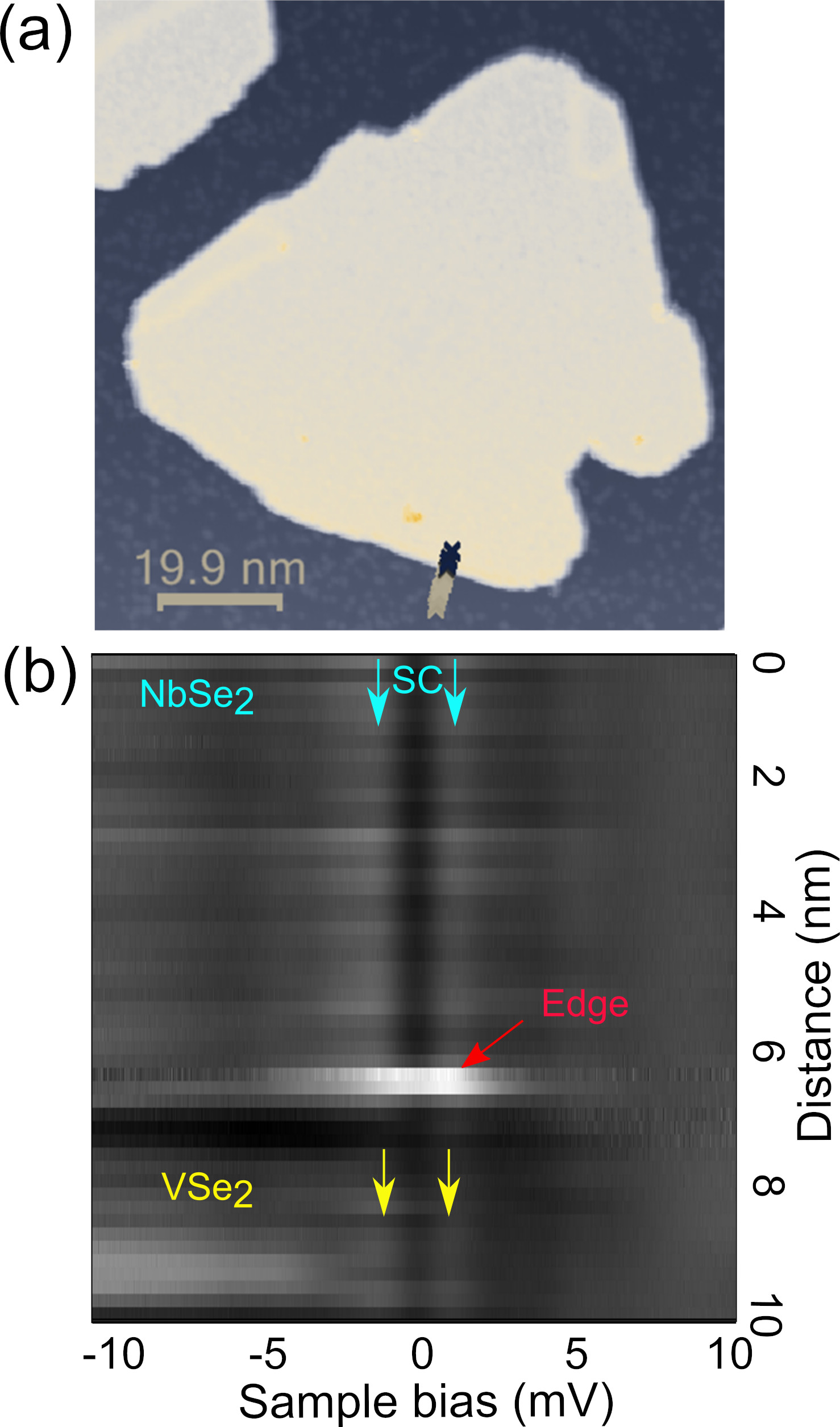}
	\caption{ localized edge mode of VSe$_2$ on NbSe$_2$. (a) STM topographic image (1 V, 10 pA) of VSe$_2$ on NbSe$_2$ (b) A line spectra taken along the edge of island where its width can be a couple of times larger than the superconducting gap width.}
	\label{FigS3}
\end{figure}
Fig.~\ref{FigS3}b shows a linecut profiles of d$I$/d$V$ spectra along the line in Fig.~\ref{FigS3}a, which cross the edge of the  VSe$_2$. As the tip approaches the island edge, the superconducting coherence peaks (white arrow in the Fig.~\ref{FigS3}b) in the d$I$/d$V$ spectra are suppressed and new electronic states emerge inside the gap as shown by the highlighted red arrow in Fig.~\ref{FigS3}b. The superconducting coherence peaks further reduces on top of the VSe$_2$. 

\bibliography{vase}